\documentclass[11pt]{article}
\usepackage{amsmath}
\usepackage{amsthm}
\usepackage{mathrsfs}
\usepackage{amsfonts}
\usepackage{amssymb}
\usepackage{stmaryrd}
\usepackage{graphicx}

\textwidth 16cm \textheight 23cm \graphicspath{{../images/}}

\date{}
\begin{document}
 \title{The weighted tunable clustering in local-world networks with incremental
 behaviors
\thanks{Supported by NSFC(No.71071090,No.10971137 and No.60873058), Department of Science and
Technology of Shandong Province(No.Z2008G04 and BS2009DX005) and STCSM (No.09XD1402500).}}
\author{ Ying-Hong Ma$^{1}$, Huijia Li$^2$,  Xiao-Dong Zhang$^{3}$ \\
\thanks{Corresponding author:\ \ \ Yinghong Ma, E-mail: \ \
yhma@sdnu.edu.cn;\ \ \ \ \ \ \   Xiao-Dong Zhang, E-mail:\ \ \ \ \ \
xiaodong@sjtu.edu.cn}
$^1$ School of Management \& Economics, Shandong Normal University,\\
        Jinan, Shandong, 250014, P.R. China.\\
$^2$  Academy of Mathematics and Systems Science, Chinese Academy of Sciences,\\
           Beijing 100080, P.R. China.\\
$^3$ Department of Mathematics, Shanghai Jiaotong  University,\\
           800 Dongchuan Road, Shanghai, 200240, P.R. China.\\
}
 \maketitle \thispagestyle{empty}

\begin{center}
\begin{minipage}{100mm}
Abstract: Since some realistic networks are influenced not only by
increment behavior but also by tunable clustering mechanism with new
nodes to be added to networks, it is interesting to characterize the
model for those actual networks. In this paper, a weighted
local-world model, which incorporates increment behavior and tunable
clustering mechanism, is proposed and its properties are
investigated, such as degree distribution and clustering
coefficient. Numerical simulations are fit to the model characters
and also display good right skewed scale-free properties.
Furthermore, the correlation of vertices in our model is studied
which shows the assortative property. Epidemic spreading process by
weighted transmission rate on the model shows that the tunable
clustering behavior has a great impact on the epidemic dynamic.


{\bf PACS classification codes:} 89.75.Hc, 05.10-a.
\end{minipage}
\end{center}

\section{Introduction}
Since Barab$\acute{a}$si and Albert introduced the classical BA
model \cite{Barab1}, lots of empirical measurements have been used
to discover some properties on real-world complex systems. However,
networks as well known are far from general scale-free structure and
people find in many real networks, for example, Internet, power
grids, WWW and so on \cite{Barab2}-\cite{Amaral}. Nodes connect each
other closer within a group than to other groups, such a group
 is defined to be a local-world which plays an important
role in micro dynamics of scale-free networks including both boolean
and weighted complex networks \cite{E. Ravasz}.

\begin{figure}[h!]
\centering
\includegraphics[angle=0, width=0.2\textwidth]{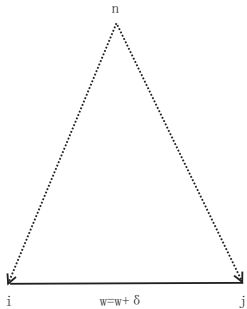}
\caption{ It is an increment behavior mechanism. That is, if a new
node $n$ connects to both the existing node $i$ and it's neighbor
$j$, then the link weight between $i$ and $j$ increases $\delta$. }
\label{Fig:1}
\end{figure}

In some complex systems, when a new individual enters the system, it
takes a great deal of searching in the global preferential
attachments. 
The preferential attachment of local economy regions exists in World
Trade Web \cite{X.Li}. In Internet on the router level, a host only
has information of others who are in a same local domain \cite{B.J.
Kim, W.X. Wang}. Motivated by the above phenomenon, Li and Chen
\cite{X.Li} presented a Local World model in which a new node makes
local preferential attachment in a local-world.  Zhang et al.
\cite{Z.Z. Zhang} introduced an evolving scale-free network model
with a continuously adjustable clustering coefficient to modify the
small cluster of Li's. Chen et al. \cite{G.R. Chen} proposed a
multi-local-world model to mimic the Internet. Wu and Liu \cite{X.
Wu} presented a high clustering coefficient in a local-world model.
On the other hand, there are many models of growing networks with
community structures, for example, Noh et al. \cite{Noh} proposed a
growing network model with group structures basing on creation or
joining mechanism.  Further, Xuan et al. \cite{Xuan} gave a
hierarchical and modular network by adding a growing rule with the
preferential attachment rule and the Motter's modeling procedure.
Pollner et al. introduced a model with the dynamics of overlapping
communities.  But these papers basically consider boolean growing
networking with local small-worlds,  while the weighted networks are
conformity with the real-world
 whose links between nodes
display heterogeneity in the capacity and intensity \cite{A.
Barrat}-\cite{Z. Pan}.
 Figure \ref{Fig:1} shows the local micro
dynamics in networks.

In previous studies, when a new node $v$ is added to the network and
builds links with an existing node $v_i$ and it's neighbor $v_j$,
the weight of link between $v_i$ and $v_j$ dose not change. However,
$v_i$ and $v_j$ is in one local-world with high probability, they
will have more common knowledge and interest if they received same
information from a node, and then the probability that they
communicate with each other will increase, that is reflexed by the
increasing of the like weight between them.    We call thus
phenomenon {\em increment behavior}, which had been noticed by
Dorogovstev et al. \cite{Dorogovstev}.  Increment behavior exists
universally in real-life networks. For example, in wide airport
networks(WAN) \cite{Guimera1,Guimera2}, the amount of passengers
between two airports will increase when they both build airlines
with a common airport, in SCN \cite{Barab3, Huberman}, two
cooperative scientists will strengthen their cooperation to a great
extent if they are collaborated with a common scientist in different
research fields. Dorogovstev et al. \cite{Dorogovstev} proposed a
growing weighted network model with a power-law weight distribution
based with the preferential attachment rule and "increment
behavior". But the communities of this  growing network model was
not discussed and studied in this paper.

In this paper, in order to character the real-life networks more
precisely, we present a weighted local-world evolving network model
with increment behavior and tunable clustering mechanism which can
capture both the dynamic of local preferential attachment(LPA)
\cite{X.Li} and triad formation(TF) \cite{P.Holme}. The  three main
mechanisms of a weighted growing networks with local small-world are
the preferential attachment of communities, strength preferential
attachment and increment behavior. It is shown that the degree of
increment behavior of this network model has a great impact on
topology property and social behaviors of networks. The geometric
characteristics of the model both analytically and numerically are
discussed, which shows that the analytical expressions are agreement
with the numerical simulations well. Moreover, we analyze the effect
of tunable clustering behaviors and increment behaviors on the
weighted local-world network. Finally, by way of the weighted
transmission rate, we  find  that the tunable clustering behavior
has a great impact on the spreading dynamic networks.

\section{The model construction}

We consider not only attachment mechanisms between local-worlds but
also inter locals preferential attachment mechanism. First, we
define three kinds of evolving attachment rules.

{\bf Local-world preferential attachment $(LPA)$:} a new node $v$ is
joining an existing local-world $\Omega_i$, the probability of
choosing $\Omega_i$, $\prod(\Omega_i)$, depends on it's size
$|\Omega_i|=N_i$, that is

\begin{equation} \label{eq:A}
\prod(\Omega_i)=\frac{N_i}{\sum_{j} N_j}.\\
\end{equation}

{\bf Strength preferential attachment $(SPA)$:} If a node $v$ had
joined in a local world, without loss of generality, $\Omega_i$,
then $v$ connected to a node $v_j$ which strength is $s_j$ in
$\Omega_i$ with probability $\prod(s_j)$,
\begin{equation} \label{eq:B}
\prod(s_j)=\frac{s_j}{\sum_{l} s_{l}}. \\
\end{equation}

For the sake of convenience, local-world preferential attachment
$(LPA)$ and strength preferential attachment $(SPA)$ are called by a
joint name {\em preferential attachment}, and a link is called {\em
PA link} if it obtained by the preferential attachment.

{\bf Increment behavior mechanism:} If a new node $v$ joins the two
nodes $v_i$ and $v_j$ of a link $v_iv_j$  in a network, then the
weight of the link $v_iv_j$
 increases $\delta$. The network growing way
according to this mechanism is illustrated in Fig. \ref{Fig:1}.

In some real-life networks, clustering structure sometime can be
quantified by large clustering coefficients, that means there are
many triangles in networks, this property will be shown in the
following construction for this model which induced by the tunable
clustering mechanism. Besides {\em PA links}, we also introduce {\em
triad-formation(TF) links} which means if a new node $v$ joins node
$v_i$ at last time, then a neighbor $v_j$ of node $v_i$ is selected
to connect $v$ with probability

\begin{equation} \label{eq:C}
\prod(w_{ij})= \frac{w_{ij}}{s_{i}}, \\
\end{equation}
where $w_{ij}$ is the weight of link $v_iv_j$. It is easy to see
that an increment behavior appears on link $v_iv_j$ with emergence
of $TF$ links.

Next, we produce our model:

{\bf Process of generation}: The weighted tunable cluster
local-world with increment behavior network model.

\begin{center}
\begin{minipage}{15cm}
Initializing an undirect weighted network with $c_0$ communities
($c_0>1$), each community is an $n_0$ complete connected network,
and there is a link for each pair communities, that is, there are
$c_0(c_0-1)/2$ inter-local links to make $c_0$ local-worlds
connected, the weight for each link $v_iv_j$ is set $w_{ij}=w_0=1$.
\\
{\bf BEGIN:}\\
 {\em step 1,} with probability $q$, a
new local-world $\Omega$ with $n_0$ completely connected nodes is
added and the weight for each link is $w_0=1$. Choose a node in
$\Omega$ randomly to connect $m$ existing nodes in other
local-worlds according to preferential attachment, it comes into $m$
$PA$ links.

{\em Step 2,} with probability $1-q$, a new node $v$ is added with
$m$ links, and the generated mechanism is preferential attachment.

{\em Step 3,} with probability $p$, a $TF$ link is introduced by
equation (\ref{eq:C}). Both $PA$ link and $TF$ link induced the
increment behavior.
END\\
\end{minipage}
\end{center}

For simply, we set the initial weight of $PA$ links and $TF$ links
with $w_0=1$ and increment amount with $\delta$ for each old link.
In the generating process, after $t$ steps, there are $c_{0}+qt$
local-worlds and $c_{0}n_{0}+(qn_{0}+(1-q))t$ nodes and
$c_{0}(n_{0}(n_{0}-1)/2+c_0(c_0-1)/2)+(q(n_{0}(n_{0}-1)/2+m(1+p))+(1-q)m(1+p))t$
links in this growing network model. So the total strength of this
network is
$2(c_{0}(n_{0}(n_{0}-1)/2+c_0(c_0-1)/2)+(q(n_{0}(n_{0}-1)/2+m(1+p+p\delta))+(1-q)m(1+p+p\delta)))t$.
And when $p=q=0$, the model is consistent with weighted BA model,
which sets the weight of edges to 1 in $BA$ model. So, at this time,
all nodes $v_i$ in this model has $k_i$=$s_i$. In the following
section, we analysis the topology properties of the weighted tunable
cluster local-world network model.

\section{Topology properties of the Model}
\subsection{Distributions of  local world on size, degree and link weight}

 By mean-field theory \cite{barabsi3}, we can obtain the
distribution of local-world size, strength, degree, and link weight
in the weighted tunable cluster local-world network model. The mean
strength $s$ is $\langle
s\rangle\approx\frac{qn_0(n_0-1)+2m(1+p+p\delta)}{qn_{0}+1-q}$, when
$p$, $q$ and $\delta$ are fixed and $t$ large enough, therefor,
$\langle s\rangle$ and $\langle w\rangle=\langle s\rangle/2$ are
considered as constants sometime.

First, considering the size distribution of local-worlds. Assume the
size of local-world is continuous, hence, $LPA$ is also interpreted
as continuous rate of change on local-world $\Omega_i$,
\begin{equation} \label{eq:C1}
\frac{\partial N_{i}}{\partial t}=\frac{N_{i}}{\sum_{j}N_j}.
\end{equation}
Then the total number of nodes in one local-world at time $t$ is
\begin{equation} \label{eq:C2}
N=\sum_{j}N_j =c_{0}n_{0}+(qn_{0}+(1-q))t\approx (qn_{0}+(1-q))t.
\end{equation}
By initial condition $N(t=t_i)=n_0$, we get
\begin{equation} \label{eq:C3}
N_{i}= n_{0}(\frac{t}{t_{i}})^{(1/1-q+qn_0)}.
\end{equation}
For convenience, we simplify probability density of $t_{i}$ by
\begin{equation} \label{eq:C4}
p(t_i)=\frac{1}{c_0+qt}.
\end{equation}
Then the size distribution of local-world $\Omega_i$ is
\begin{equation} \label{eq:C5}
\begin{split}
P(N_i(t)\geq n)=P(t_{i}\leq
\frac{n_{0}^{1-q+qn_{0}}}{n^{1-q+qn_{0}}}t)
=\frac{1}{c_{0}+qt}(\frac{n_0}{n})^{1+q(n_0-1)}.
\end{split}
\end{equation}
Equation (\ref{eq:C5}) shows that the size distribution of
$\Omega_i$ obeys power-law rule, $P(N)\sim N^{-\gamma}$ and
$\gamma=1+q(n_{0}-1)\geq 1$. This property is in conformity with
many real-world networks
   \cite{Barab2,B.A. Huberman, M.E.J. Newman, Z. Pan,  Dorogovstev,Guimera1}, which are scale-free
local-world with scaling exponent $\gamma\in [1,2]$.

Next, we analyze the strength distribution, the degree distribution,
and the weight distribution.

\begin{figure}[h!]
\centering
\includegraphics[angle=0, width=0.13\textwidth]{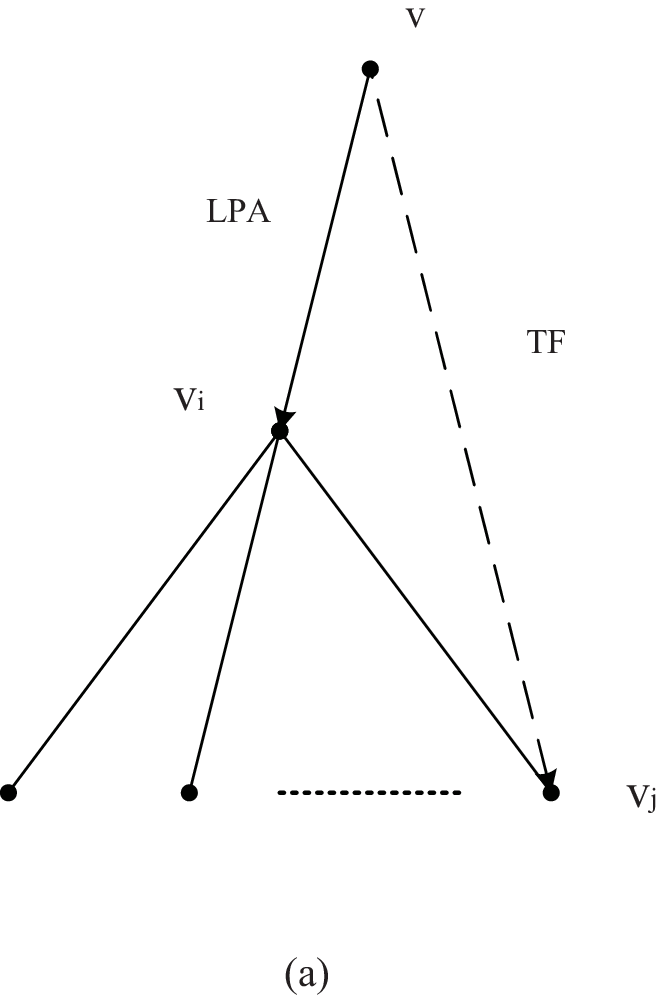}
\includegraphics[angle=0, width=0.14\textwidth]{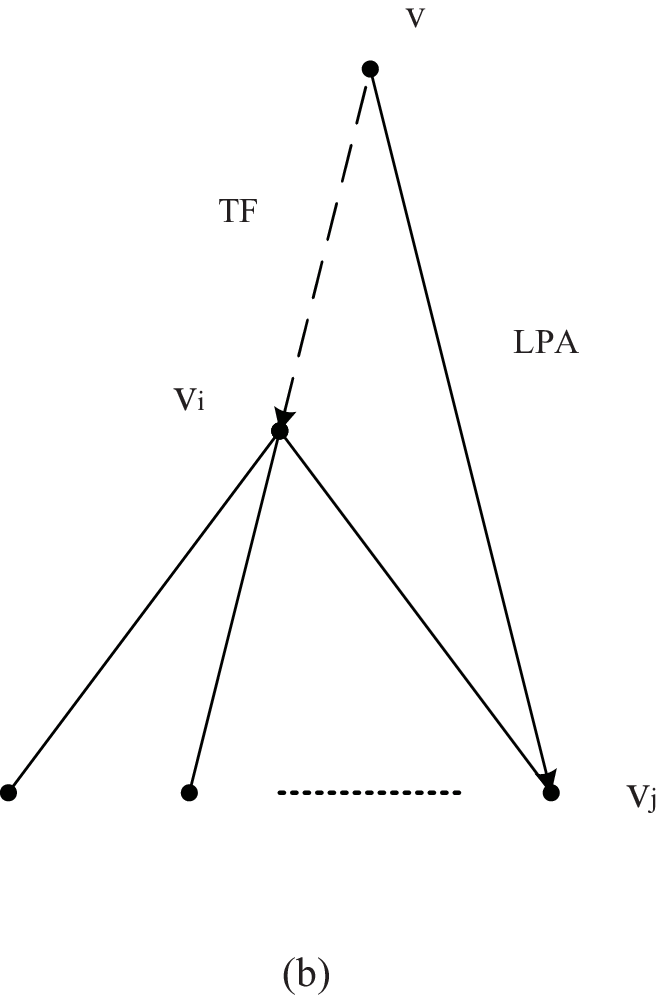}
\includegraphics[angle=0, width=0.125\textwidth]{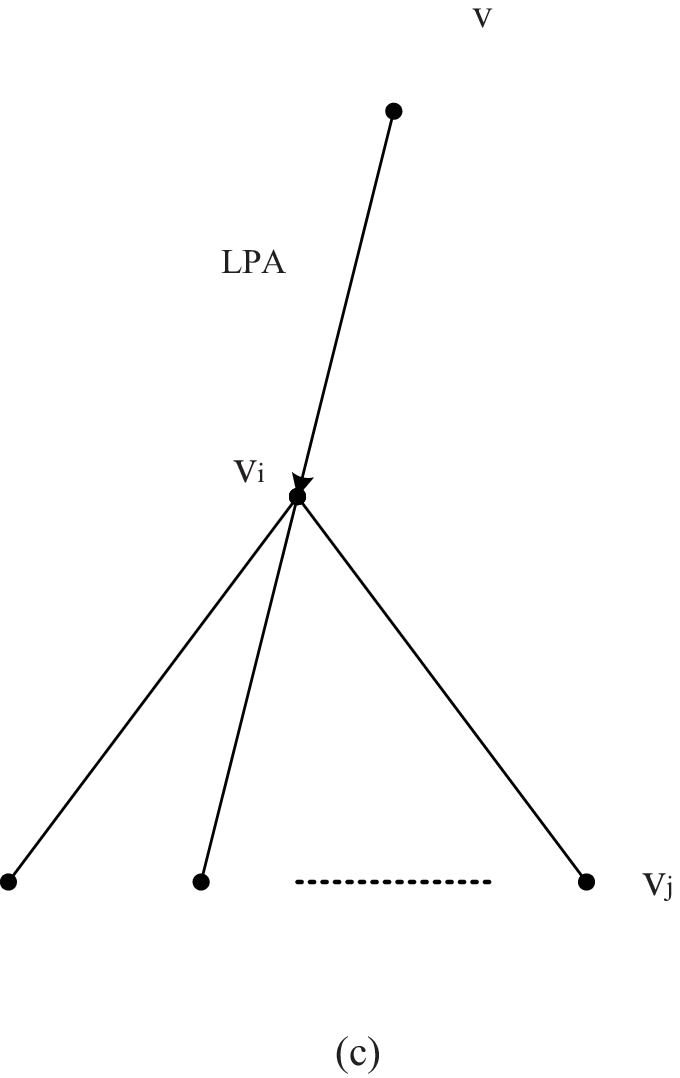}
\caption{The microscopic variety mechanisms for $s_i$. The dashed
edges are $TF$ links.(a) $v$ joins $v_i$ by $LPA$, a neighbor $v_j$
of $v_i$ is chosen to build a $TF$ link. In this case,
$s_i=s_i+1+\delta$; (b) $v$ connects to $v_j$ which is a neighbor of
$v_i$ by $SPA$, and $v_iv$ is a $TF$ link. In this case,
$s_i=s_i+1+\delta$. (c) $v$ joins $v_i$ by $SPA$ but none of $v_i$'s
neighbor is selected to build $TF$ link with $v$. In this case,
$s_i=s_i+1$.} \label{Fig:2}
\end{figure}

If a new node $v$ is selected by $SPA$ and added to a local-world,
the strength $s_{i}$ of an old node $v_i$ will be affected by $v$ in
following three cases, those microscopic variety mechanisms for
$s_i$ are shown in Figure \ref{Fig:2}.

By Figure \ref{Fig:2}, it is easy to see that the strength change
rate of $v_i$ is

\begin{equation} \label{eq:C6}
\begin{split}
\partial s_{i}/\partial &t=
m\frac{N_{i}}{\sum{N_{j}}}\frac{s_{i}}{\sum_{v_l\in
local}s_l}(\sum_{v_l\in \Omega_i}\frac{w_{il}}{s_{i}}p(1+\delta)+
\sum_{v_l\in
\Omega_i}\frac{w_{il}}{s_{l}}p(1+\delta)+\sum_{j\in \Omega_i}\frac{w_{ij}}{s_i}(1-p))\\
&=m\frac{N_{i}}{\sum{N_j}}\frac{s_i}{\sum_{v_l\in
local}s_l}(1+p+2p\delta),
\end{split}
\end{equation}
 and after $t$ steps, the total strength of each local-world in model, on average,
 is

\begin{equation} \label{eq:C7}
\begin{split}
 \sum_{v_j\in local} s_j=\langle
s\rangle N_{i}.
\end{split}
\end{equation}
From Figure \ref{Fig:2}, the degree of $v_i$ is affected in cases
(a) and (b). However, increment behavior does not have any impact in
case (c). Therefor, degree change rate of $v_i$ is
\begin{equation} \label{eq:C8}
\begin{split}
\partial k_{i}/\partial t
&= m\frac{N_{i}}{\sum{N_{j}}}(\frac{s_{i}}{\sum_{v_l\in local}s_l}+
\sum_{v_j\in \Omega_i}{\frac{s_j}{\sum_{v_l\in
local}s_l}}\frac{w_{ij}}{s_{j}}p)\\
&=m\frac{N_{i}}{\sum{N_{j}}}\frac{s_i}{\sum_{v_j\in local}s_j}(1+p).
\end{split}
\end{equation}
When $t$ is large enough and after $t$ steps, the amount of nodes in
any local-world is
\begin{equation} \label{eq:C9}
\begin{split}
\sum_{j}N_j=c_{0}n_{0}+(qn_0+(1-q))\approx (qn_{0}+(1-q))t.
\end{split}
\end{equation}
Hence, we obtain
\begin{equation} \label{eq:C10}
\begin{split}
\partial s_{i}/\partial t=\frac{m(1+p+2p\delta)}{\langle
s\rangle(qn_{0}+1-q)}\frac{s_{i}}{t} =A\frac{s_{i}}{t},
\end{split}
\end{equation}
where $A=m(1+p+2p\delta)/(\langle s\rangle(qn_{0}+1-q))$. And
\begin{equation} \label{eq:C11}
\begin{split}
\partial k_{i}/\partial t=\frac{m(1+p)}{\langle
s\rangle(qn_{0}+1-q)}\frac{s_{i}}{t} =C\frac{s_{i}}{t},
\end{split}
\end{equation}
where $C=m(1+p)/(\langle s\rangle(qn_{0}+1-q))$.
 With the initial time $t_i$ of $v_i$ ,
$s_i(t_i)=k_i(t_i)=m(1+p)$ and  equations (\ref{eq:C10})
(\ref{eq:C11}), we have
\begin{equation} \label{eq:C12}
\begin{split}
s_i(t)=m(1+p)(\frac{t}{t_{i}})^{A},
\end{split}
\end{equation}
\begin{equation} \label{eq:C13}
\begin{split}
k_i(t)=\frac{s_i(t)+(A/C-1)m(1+p)}{A/C}.
\end{split}
\end{equation}
Therefore, by mean-field theory, the strength distribution is
\begin{equation} \label{eq:C14}
\begin{split}
p(s)=\frac{\partial P(s_i(t)<s)}{\partial s}
=\frac{1}{c_0n_0+(qn_0+(1-q))}\frac{m(1+p)^{1/A}}{A} s^{-(1+1/A)},
\end{split}
\end{equation}
since the time density is $p(t=t_i)=\frac{1}{c_0n_0+(qn_0+(1-q))t}$.
By equation (\ref{eq:C13}), clearly, it is a linear relationship
between strength and degree of $v_i$. Therefore, the degree
distributions have the same distribution with strength, hence
 $P(k)\sim k^{-\gamma}$ and $\gamma=1+1/A$.

When a new node $v$ connects two ends of a link $v_iv_j$,  the link
weight $w_{ij}$ will change by (a) and (b) in Figure \ref{Fig:2},
that is,
\begin{equation} \label{eq:C15}
\begin{split}
\partial w_{ij}/\partial t &=
m\frac{N_{i}}{\sum{N_{j}}}(\frac{s_i}{\sum_{v_l\in
local}s_l}p\frac{w_{ij}}{s_i}\delta+\frac{s_j}{\sum_{v_l\in
local}s_l}p\frac{w_{ij}}{s_j}\delta)\\
&=\frac{2mp\delta}{\langle
s\rangle(qn_{0}+1-q)}\frac{w_{ij}}{t}=D\frac{w_{ij}}{t},
\end{split}
\end{equation}
where $D=(2mp\delta)/(\langle s\rangle(qn_{0}+1-q))$. Therefore,
$w_{ij}=(\frac{t}{t_i})^D$ with initial time $t_i$ of $v_i$ and
$w_{ij}(t_{i})$=$1$. So  the density of the link weight distribution
is
\begin{equation} \label{eq:C16}
\begin{split}
p(w)=\frac{\partial P(w_{ij}(t)<w)}{\partial w}
=\frac{1}{c_0n_0+(qn_0+(1-q))}\frac{1}{D} w^{-(1+1/D)}.
\end{split}
\end{equation}
\begin{figure}[h]
\centering
\includegraphics[angle=0, width=0.4\textwidth]{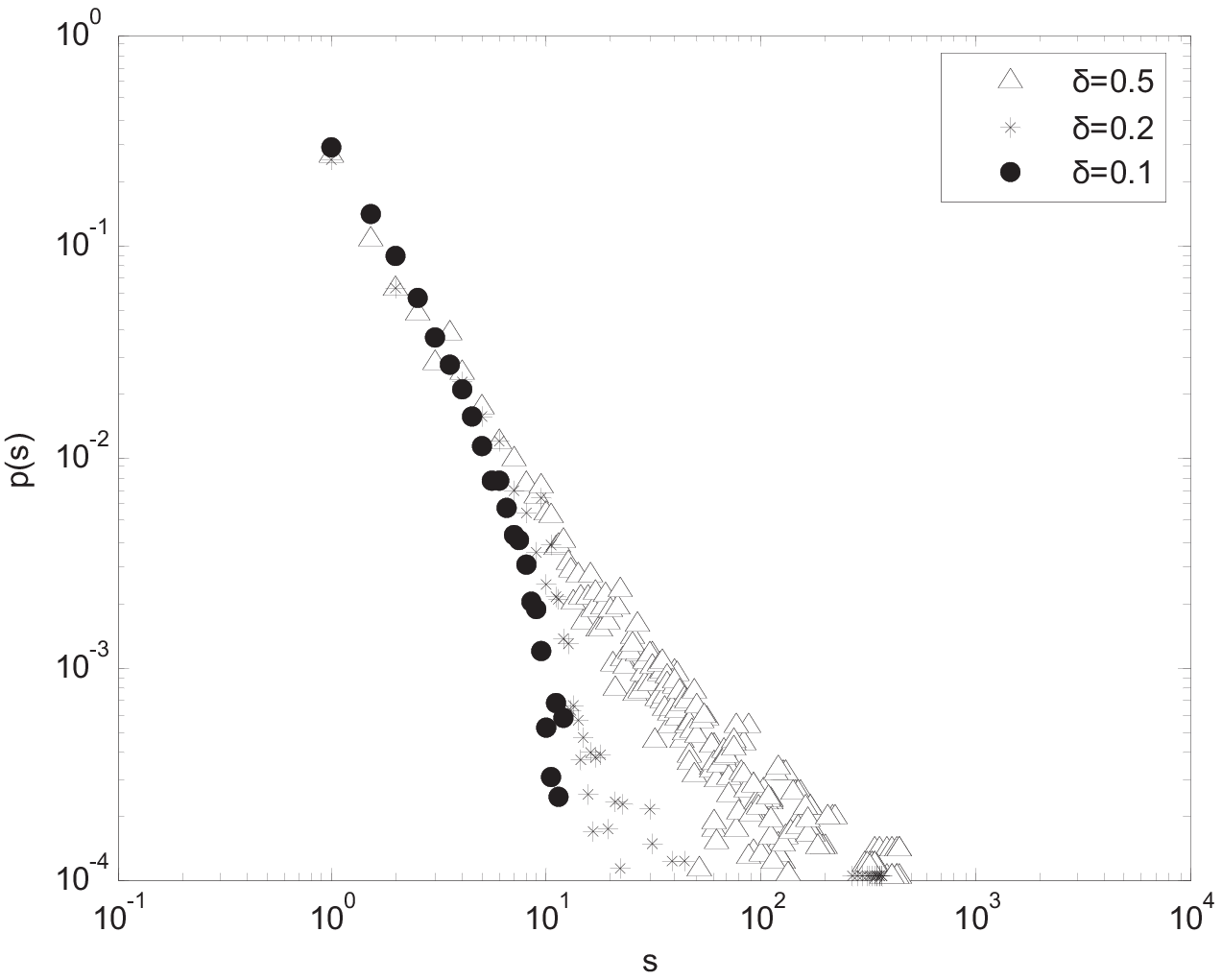}
\includegraphics[angle=0, width=0.4\textwidth]{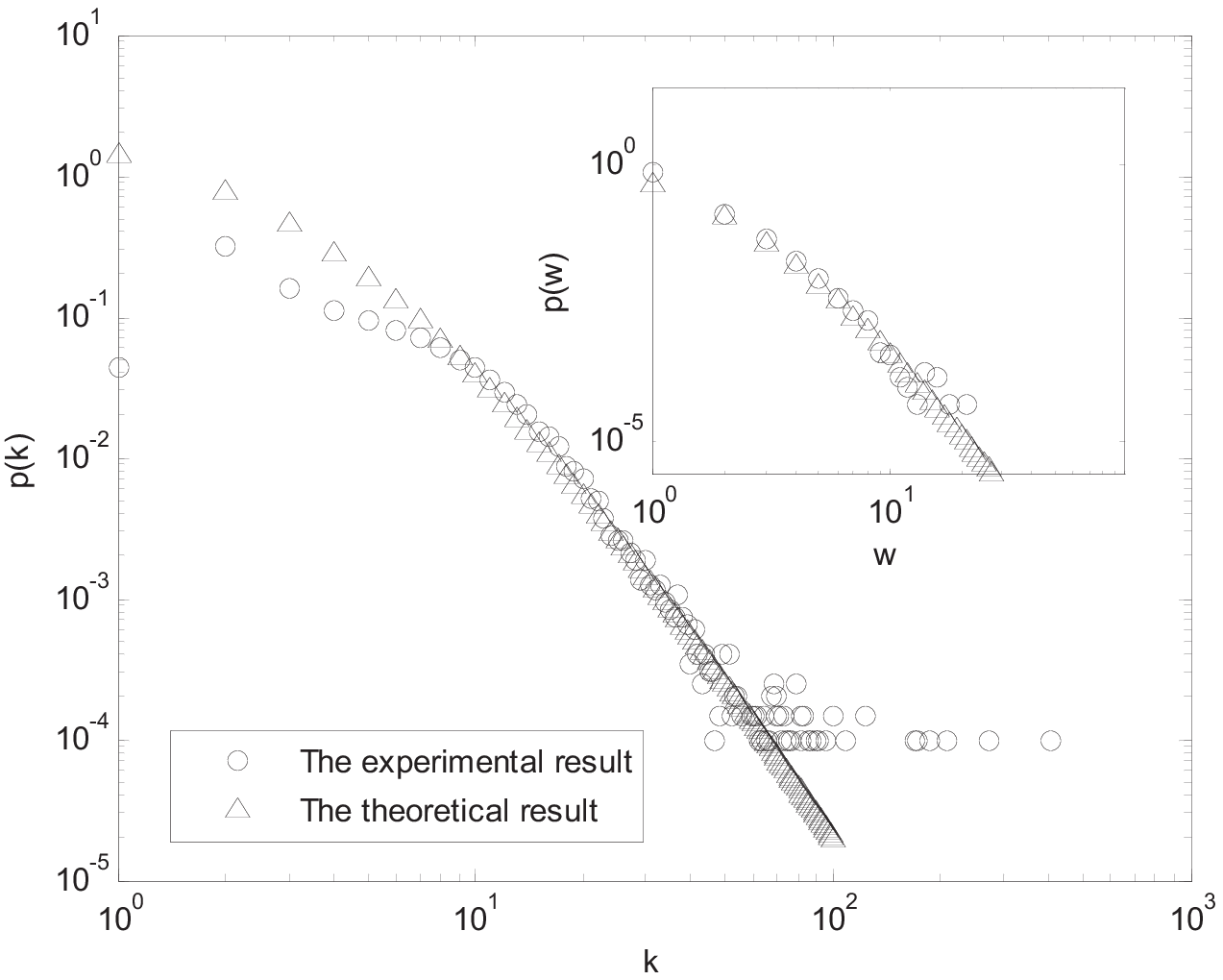}
\caption{ The two graphs show the strength distributions. The left
graph shows the status when $m=3,\, p=0.1,\, q=0.01$ and
$\delta=0.1,\, 0.2,\,0.5$ in a log-log plot, where $p(s)$ is the
probability density of strength $s$. The right graph illustrates the
degree distributions with $m=3, p=0.1$ and $q=0.01$ and the inner
graph is the state of the link weight distribution $p(w)$. All data
are realized in averaging over 50 independent simulations with fixed
size $N=10^4$.} \label{Fig:Cl}
\end{figure}
The left graph in Fig.\ref{Fig:Cl} displays the strength
distribution on three values of $\delta$ and the right graph
exhibits the degree distribution and the weight distribution in a
special case with theoretical curves. All those simulations agree
well with Eqs.~(\ref{eq:C14}) and (\ref{eq:C16}) obtained from
theory. It is observed that there is some deviation from power-law
behavior alone with small values of $s$ and $k$, which are
originated by fluctuation on the number of new links built-up by
systems (\cite{ZLiu1, ZLiu2}). In many real-life networks, such as
the World Wide Web \cite{Caldarelli, Amaral}, the actor
collaboration network \cite{D. J. Watts} and scientific
collaboration networks \cite{Barab3,Huberman}, truly exhibit those
phenomenons of deviation from power-law in small $s$ and $k$ values.

\subsection{Generalization of clustering on the weighted networks}

Large clustering coefficient is a typical property of social
acquaintance networks, where two individuals with a common friend
are likely to know each other. The clustering coefficient of a node
indicates the close level among its neighborhood. The clustering
coefficient $c_i$ of a node $v_i$ is the ratio of the existed link
size $e_i$ and all possible links size among its $n_i$ neighbors,
that is, $c_i=2e_i/(n_i(n_i-1))$. The clustering coefficient of a
network is the average for all individual $c_i$. In order to
understand the structure organization of weighted networks much
better, the weighted clustering coefficient $c_i^w$ is introduced by
Barrat et al. \cite{Barrat}.
\begin{equation} \label{eq:D1}
 c_i^w=\frac{1}{s_i(n_i-1)}\sum_{j,h}\frac{(w_{ij}+w_{ih})}{2}a_{jh},\\
\end{equation}
where $s_i$ is the strength of $v_i$, $w_{ij}$ is the weight of link
$v_iv_j$ and $a_{jh}$ is the element of adjacent matrix in which
$a_{jh}=1$ if $v_jv_h$ is a link,  $a_{jh}= 0$ otherwise. In here,
the weight for each link is positive. If  weights of some links are
zero, then the topological structures of the weighted networks will
be different. Therefore, the weighted clustering coefficient without
an edge is not the same as if the weight of the link is put to zero.
The clustering coefficient of weighted networks depends on both its
topological structure and its weight of each link.  However, Onnela
et al. in \cite{onnela2005} and Holme in \cite{holme2007} provided a
different definition of the clustering coefficient of weighted
networks, which is only dependent on its weight, while independent
of its topological structure. For the sake of stressing known
topological structure of networks, we adapt the definition of the
clustering coefficient of weighted networks in \cite{Barrat}.

\begin{figure}[h!]
\centering
\includegraphics[angle=0, width=0.12\textwidth]{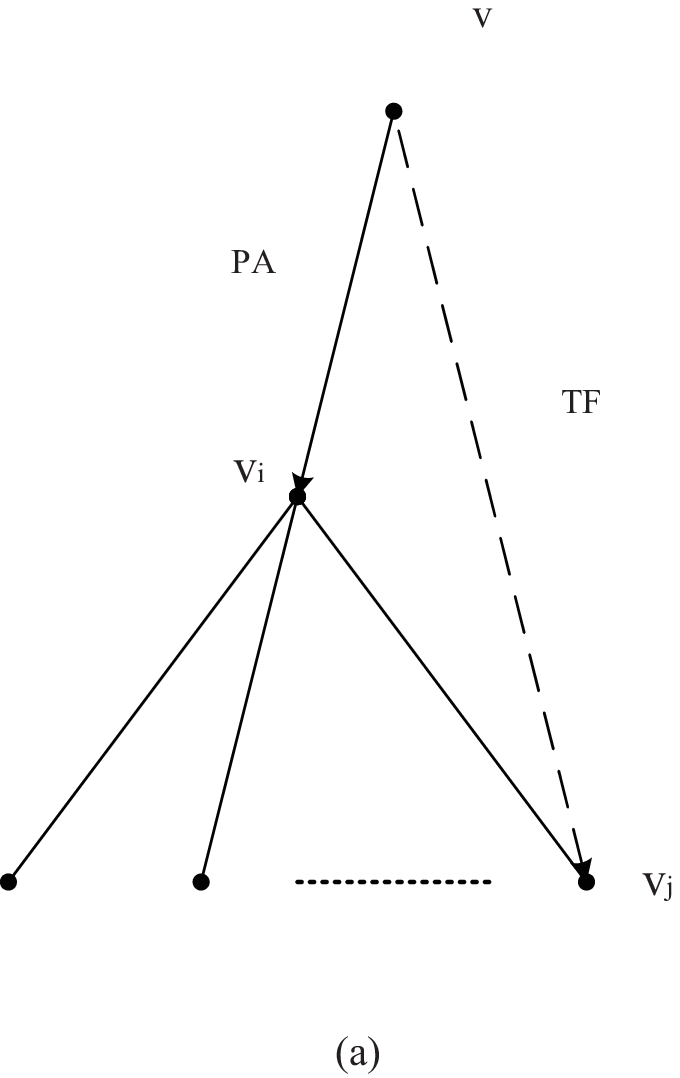}
\includegraphics[angle=0, width=0.12\textwidth]{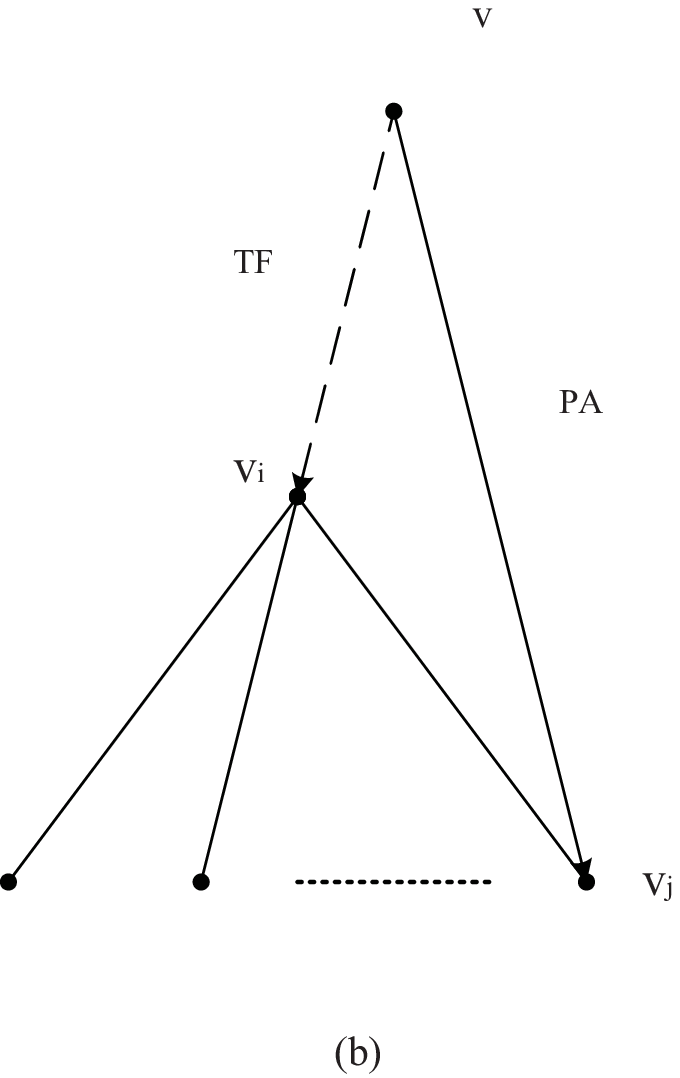}
\includegraphics[angle=0, width=0.12\textwidth]{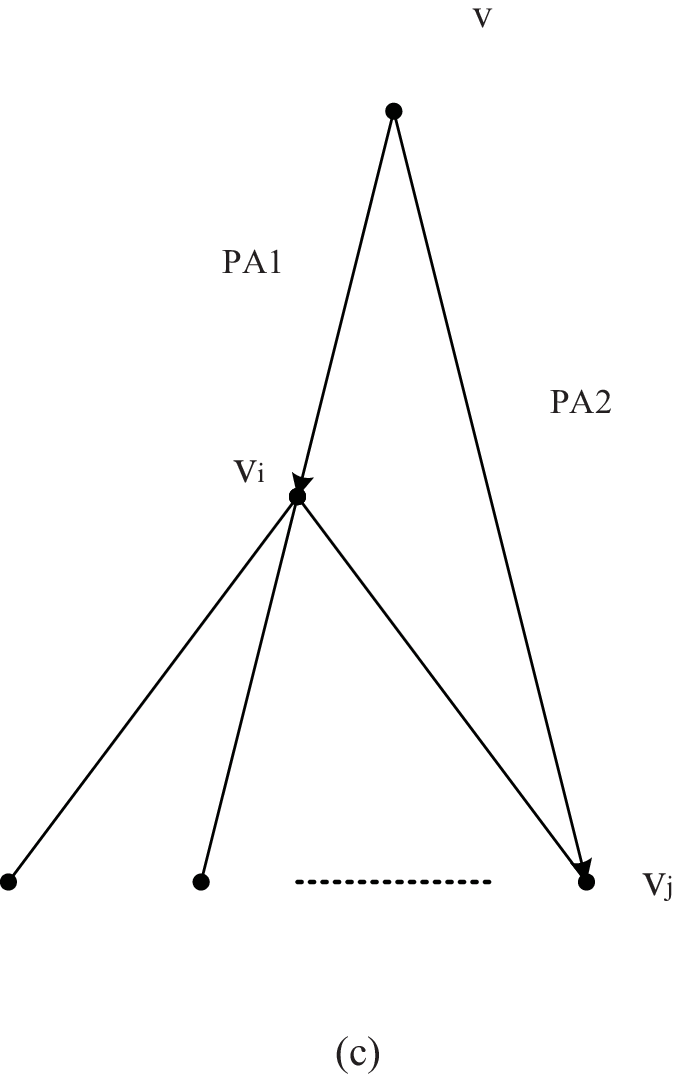}
\includegraphics[angle=0, width=0.12\textwidth]{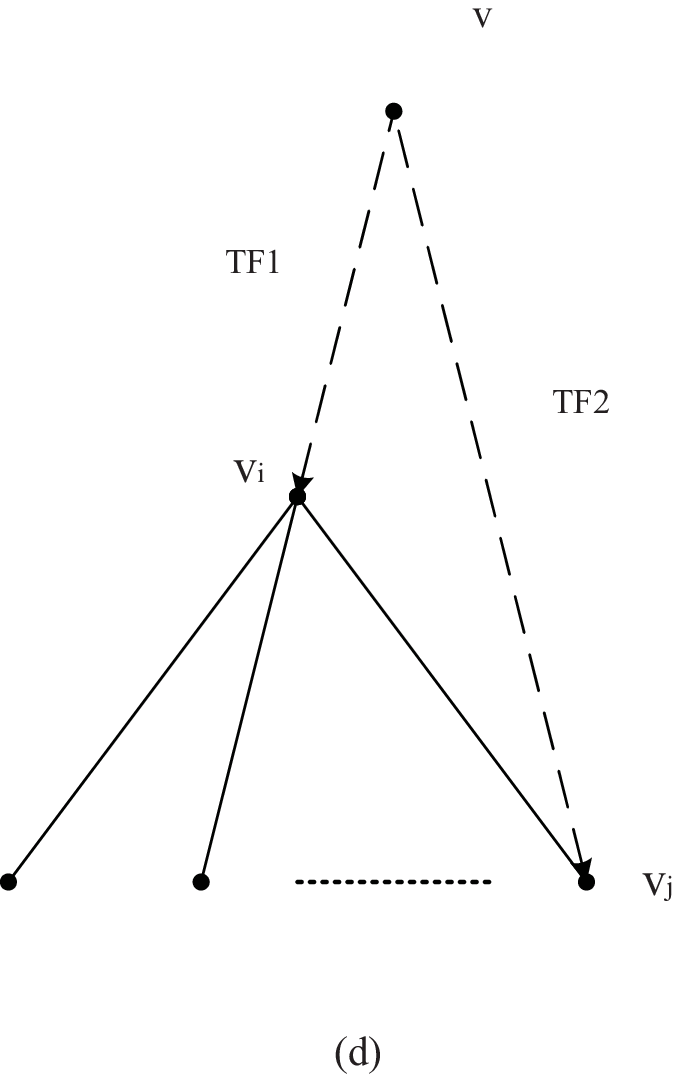}
\includegraphics[angle=0, width=0.12\textwidth]{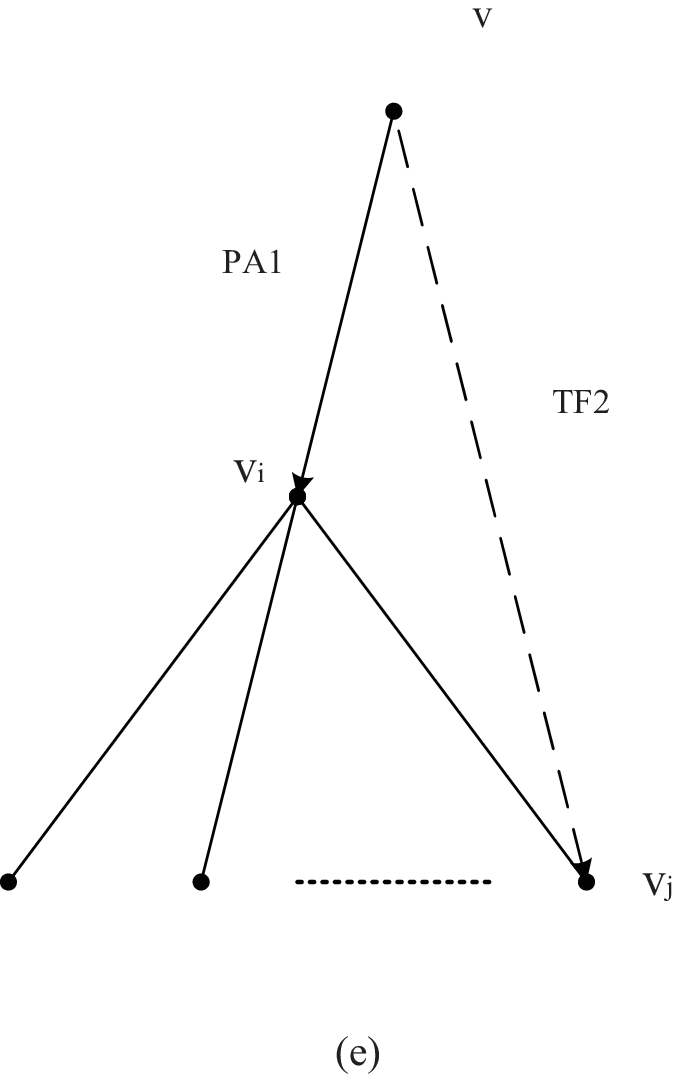}
\includegraphics[angle=0, width=0.12\textwidth]{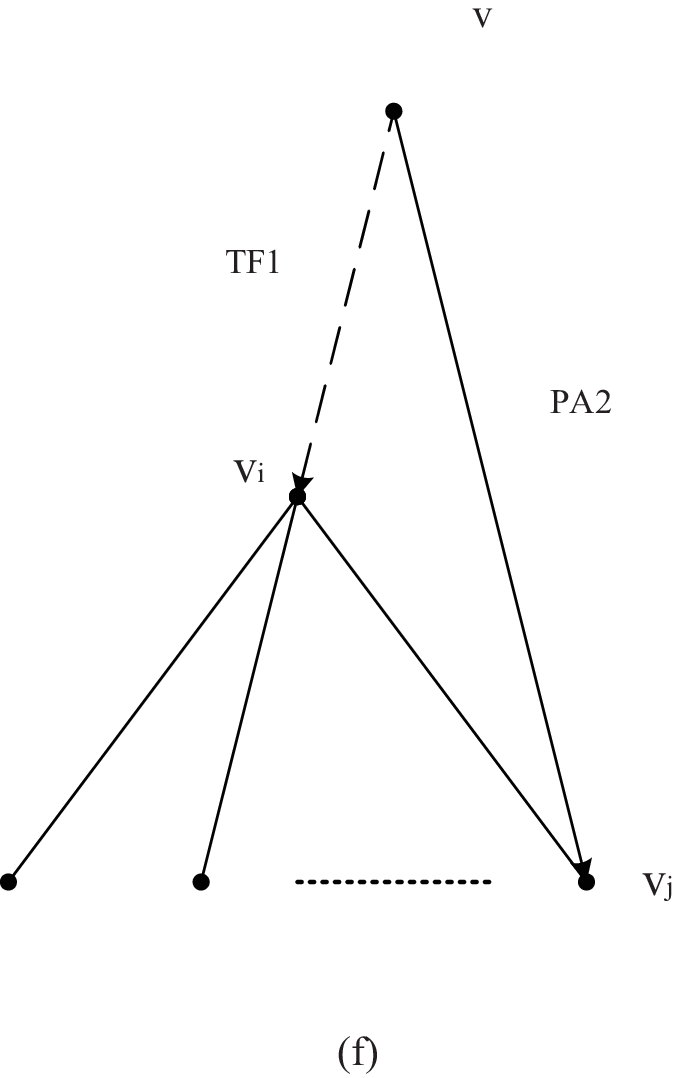}
\caption{The dashed links show the change microscopic mechanisms of
$e_i$. (a) $v_i$ connects to $v$ by $PA$ step, and potentially
following by a $TF$ step which successes $vv_j$ be a link. (b) By a
$PA$ step, $v$ attaches to $v_j$ which is one of neighbor of $v_i$,
and by one of $TF$ step $v$ conversely links to $v_i$. (c) $v_i$
connects to $v$ by a $PA$ step, $v_j$, a neighbor of $v_i$, in
another $PA$ step is selected to connect $v$. (d) By a $TF$ step,
$v_i$ connects to $v$, and by another $TF$ step, $v$ joins $v_j$, a
neighbor of $v_i$. (e) $v$ connects to $v_i$ by a $PA$ step, and
following by a potential $TF$ step such that $v$ connects to $v_j$,
a neighbor of $v_i$. (f) $v_i$ connects to $v$ by a $TF$ step, and
in a following potential $PA$ step, $v$ links to $v_j$.}
\label{Fig:3}
\end{figure}

The weighted clustering coefficient is only related to
the weights of links adjacent to $v_i$ and does not reflet the relations among
the neighbors. So, we extend $c_i^w$ to $\tilde{c}_i^w$ which took
all links weight among $v_i$'s neighbors and $v_i$ into account.

$$\tilde{c}_i^w= \frac{1}{s_i(k_i-1)}\sum_{j,h}\frac{(w_{ij}+w_{ih}+w_{jh})}{3}
a_{jh}=\frac{\langle w_{ijh}\rangle}{s_i(k_i-1)}e_i,$$ where
$\langle w_{ijh}\rangle$ is the average weight of all the new
triangles $v_iv_jv_h$ with $v_i$ newly added to networks. In all the
new triangles $v_iv_jv_h$, the edge weight $w_{ij}=w_{ih}=1$ since
$v_iv_j$ and $v_iv_h$ are new edges,  and $w_{jh}=w_{jh}+\delta$
because $v_jv_h$ is an old edge with the increment behavior arising
when the new node $v_i$ is joined $v_j$ and $v_h$. So  $\langle
w_{ijh}\rangle$ can be approximated by
$\frac{1}{3}(2+(\frac{t}{t_{i}})^{D}+\delta)$ with
$w_{ij}=(\frac{t}{t_i})^D$. Therefore,
\begin{equation} \label{eq:D2}
\begin{split}
&\tilde{c}_i^w=\frac{2+(\frac{t}{t_{i}})^{D}+\delta}{s_i(k_i-1)}e_i.
\end{split}
\end{equation}
In equation (\ref{eq:D2}), the main parameter is links size $e_i$,
in Fig.\ref{Fig:3} illustrates microscopic change mechanisms of
$e_i$, where $v$ is a new node selected by $LPA$ and added to a
local-world, there are six microscopic variety mechanisms showing in
Fig.\ref{Fig:3}. We can compute the variety rate of $e_i$ in Figures
\ref{Fig:3} and \ref{Fig:2}.  Therefore,

\begin{equation} \label{eq:D3}
\begin{split}
&\partial e_{i}/\partial t\\
&=m\frac{N_i}{\sum{N_j}}(\frac{s_{i}}{\sum_{v_j\in
local}s_j}\sum_{v_k\in \Gamma_i}\frac{w_{ik}}{s_{i}}p+ \sum_{v_k\in
\Gamma_i}\frac{s_{k}}{\sum_{v_j\in
local}s_j}\frac{w_{ik}}{s_{k}}p\\
&+\frac{s_{i}}{\sum_{v_j\in local}s_j}(m-1)\sum_{v_j\in
\Gamma_i}\frac{N_{i}}{\sum{N_{j}}}\frac{s_j}{\sum_{v_k\in
local}s_k}\\
&+\sum_{v_k\in \Gamma_i}\frac{s_k}{\sum_{v_j\in
local}s_j}\frac{w_{ki}}{s_{k}}p(m-1)\sum_{v_i\in
\Gamma_i}\frac{N_{i}}{\sum{N_{j}}}\sum_{v_k\in
\Gamma_i}\frac{s_k}{\sum_{v_j\in
local}s_j}\frac{w_{ki}}{s_{k}}p\\
&+\frac{s_i}{\sum_{v_j\in local}s_j}(m-1)\sum_{v_i\in
\Gamma_i}\frac{N_{i}}{\sum{N_{j}}}\sum_{v_k\in
\Gamma_i}\frac{s_{k}}{\sum_{v_j\in
local}s_j}\frac{w_{ki}}{s_{k}}p\\
&+\sum_{v_k\in \Gamma_i}\frac{s_{i}}{\sum_{v_j\in
local}s_j}\frac{w_{ik}}{s_{i}}p(m-1)\sum_{v_j\in
\Gamma_i}\frac{N_{i}}{\sum{N_{j}}}\frac{s_{j}}{\sum_{v_k\in
local}s_k})\\
&=2pm\frac{N_i}{\sum{N_{j}}}\frac{s_{i}}{\sum_{v_j\in local}s_j}+
m(m-1)(1+p)^2(\frac{N_{i}}{\sum{N_{j}}})^2\frac{s_{i}}{\sum_{v_j\in
local}s_j}\sum_{v_k\in \Gamma_i}\frac{s_{k}}{\sum_{v_j\in
local}s_j}\\
&=\Delta_1+\Delta_2,
\end{split}
\end{equation}
where $\Gamma_i$ is the neighborhood of $v_i$.

As at the initial time $t=t_{i}$ of $v_i$, $s_i(t=t_i)=m(1+p)$,  and
by  equations (\ref{eq:C2}), (\ref{eq:C7}) and (\ref{eq:C12}), we
calculate $e_i$ by integrating both sides in (\ref{eq:D3}).

The first part in the right hand of (\ref{eq:D3}) is integrated as

\begin{equation} \label{eq:D4}
\begin{split}
\Delta_1=\int^{t_N}_{t_i}\frac{2mp}{\langle
s\rangle(qn_0+1-q)}\frac{s_i(t)}{t}\partial t
=\frac{2mp}{A\langle
s\rangle(qn_0+1-q)}(s_i(t_N)-s_{i}(t_i)).
\end{split}
\end{equation}
In order to obtain the second term $\Delta_2$ in equation
(\ref{eq:D3}), first, we consider the weighted tunable cluster
local-world model in section 2. The degree of node $v_i$ is given by
(\ref{eq:C13}) at time $t$, and with equation (\ref{eq:C12}), and
neglecting the constant part, then $k_i(t)$ is
\begin{equation} \label{eq:D7}
\begin{split}
 &k_{i}(t)\approx\frac{C}{A}s_{i}(t)
 =\frac{C}{A}m(1+p)(\frac{t}{t_{i}})^A.
\end{split}
\end{equation}
Clearly the density of time is
\begin{equation} \label{eq:D8}
\begin{split}
 &p_{i}(t)=\frac{1}{c_{0}n_{0}+(qn_{0}+1-q)t}
=\frac{1}{N}
\end{split}
\end{equation}
by equation  (\ref{eq:C2}). Then the degree of $v_i$ can also be
expressed by $k_{i}(N)=\frac{C}{A}m(1+p)(\frac{N}{i})^A$.

Second, we consider the neighborhood information of nodes in our
network model, the average degree of nodes in $\Gamma_i$ is
\begin{equation} \label{eq:D5}
\begin{split}
k_i^{nn}=\frac{1}{k_i}\sum_{v_j\in \Gamma_i}k_j.
\end{split}
\end{equation}
Let $p_{c}(k'|k)$ be conditional probability such that a link
adjacent a node with degree $k$ to a node with degree $k'$, then
equation (\ref{eq:D5}) is
\begin{equation} \label{eq:D6}
\begin{split}
\langle k^{nn}\rangle=\sum_{k'}k'p_{c}(k'|k)
=\sum_{k'}\frac{(k')^2}{\langle k\rangle}p(k') =\frac{\langle
k^2\rangle}{\langle k\rangle},
\end{split}
\end{equation}
where $p_{c}(k'|k)=\frac{k'p(k')}{\langle k\rangle}$ in uncorrelated
network. Therefore, by  (\ref{eq:D7}), (\ref{eq:D8}) and
(\ref{eq:D6}), we obtain

\begin{equation} \label{eq:D9}
\begin{split}
\langle k^2\rangle=\sum_{i=1}^{N}k_{i}^{2}(t)p_{i}(t)
=\left(\frac{Cm(1+p)}{A}\right)^2\ \ \frac{1-N^{2A-1}}{-2A+1}.\\
\end{split}
\end{equation}
Hence, the degree sum of all nodes in $\Gamma_i$ is followed by
equations (\ref{eq:D9}) and (\ref{eq:D6})
\begin{equation} \label{eq:D10}
\begin{split}
&\sum_{v_j\in \Gamma_i}k_{j}(t)=k_{i}\langle k^{nn}\rangle
=\left(\frac{C}{A}\right)^2\frac{(m(1+p))^2}{\langle k\rangle}\ \ \frac{N^{-2A+1}-1}{-2A+1}k_{i}(t).\\
\end{split}
\end{equation}
Consequently, the strength sum of all nodes in neighborhood $\Gamma_i$
of $v_i$ is
\begin{equation} \label{eq:D11}
\begin{split}
&\sum_{v_j\in \Gamma_i}s_{j}(t)
=U\frac{N^{-2A+1}-1}{-2A+1}s_{i}(t),\\
\end{split}
\end{equation}
since $k_{i}(t)$ and $s_{i}(t)$ are linear related, where
$U=(\frac{C}{A})^2\frac{(m(1+p))^2}{\langle k\rangle}$ and
$-2A+1\not=0$.


{\em Case 1.} $-2A+1<0$. That is, $A>1/2$. Then
$\frac{N^{-2A+1}-1}{-2A+1}\approx\frac{1}{2A-1}$ with $N\rightarrow
\infty$, therefore, equation (\ref{eq:D11}) is approximately
$\sum_{j\in Nei_i}s_j(t)\approx\frac{U}{2A-1}s_i(t)$.

In this case, the second term of (\ref{eq:D3}) is
\begin{equation} \label{eq:D12}
\begin{split}
&\Delta_2=\int^{t_{N}}_{t_{i}}(m)(m-1)(1+p)^2\frac{U}{2A-1}\frac{s_i(t)^{2}}{(\sum{N_i)^{2}}\langle
s\rangle^2}\partial t\\
&=\frac{Um(m-1)(1+p)^2}{\langle
s\rangle^{2}(qn_0+1-q)^2(2A-1)}\int^{t_N}_{t_i}\frac{s_i^2(t)}{t^2}\partial t\\
&=E_1\int^{t_N}_{t_i}\frac{(m(1+p))^2t^{2A-2}}{t_i^{2A}}\partial t\\
&=\frac{E_1(m(1+p))^2}{t_i^{2A}}\cdot\frac{t^{2A-1}}{2A-1}|^{t_N}_{t_i},\\
\end{split}
\end{equation}
where $E_1=\frac{Um(m-1)(1+p)^2}{\langle
s\rangle^{2}(qn_0+1-q)^2(2A-1)}$. Hence, by equations (\ref{eq:D4})
and (\ref{eq:D12}), we can get $e_i$ from equation (\ref{eq:D3}),
\begin{equation} \label{eq:D13}
\begin{split}
e_{i}&=\Delta_1+\Delta_2\\
&=\frac{2mp}{A\langle
s\rangle(qn_{0}+1-q)}(s_{i}(t_{N})-s_{i}(t_{i}))
+\frac{E_{1}(m(1+p))^2}{t_i^{2A}}\cdot\frac{t^{2A-1}}{2A-1}|^{t_N}_{t_{i}}\\
&=\Delta_0+\frac{2mp}{A\langle
s\rangle(qn_{0}+1-q)}s_{i}(t_{N})+\frac{E_{1}}{2A-1}\frac{s_i(t_N)^{2}}{t_N}\\
&\approx\frac{2mp}{A\langle
s\rangle(qn_{0}+1-q)}s_{i}(t_{N})+\frac{E_{1}}{2A-1}\frac{s_i(t_N)^{2}}{t_N},
\end{split}
\end{equation}
where $\Delta_0=-\frac{2mp}{A\langle
s\rangle(qn_{0}+1-q)}s_i(t_i)-\frac{E_1}{2A-1}\frac{s_i(t_i)^{2}}{t_i}$
is a constant, since $t_i$ is the initial time and $\Delta_0$ can
be neglected. By the above analysis, the weighted clustering
coefficient of $v_i$ with large strength $s_i(t_N)$ is obtained by
the following,
\begin{equation} \label{eq:D14}
\begin{split}
\tilde{c}_i^w&=\frac{\langle w_{ijh}\rangle}{s_i(t_N)(k_i(t_N)-1)}e_{i}\\
&\approx \frac{\langle w_{ijh}\rangle}{s_i(t_N)(k_i(t_N)-1)}(\frac{2mp}{A\langle
s\rangle(qn_{0}+1-q)}s_{i}(t_N)+\frac{E_{1}}{2A-1}\frac{s_i^2(t_N)}{t_N})\\
&\approx \frac{2mp\langle w_{ijh}\rangle}{C\langle
s\rangle(qn_0+1-q)}\frac{1}{s_i(t_N)}+\frac{E_1\langle
w_{ijh}\rangle}{2A-1}\frac{A}{C}\frac{1}{t_N}.
\end{split}
\end{equation}

\begin{figure}[h]
\centering
\includegraphics[angle=0, width=0.4\textwidth]{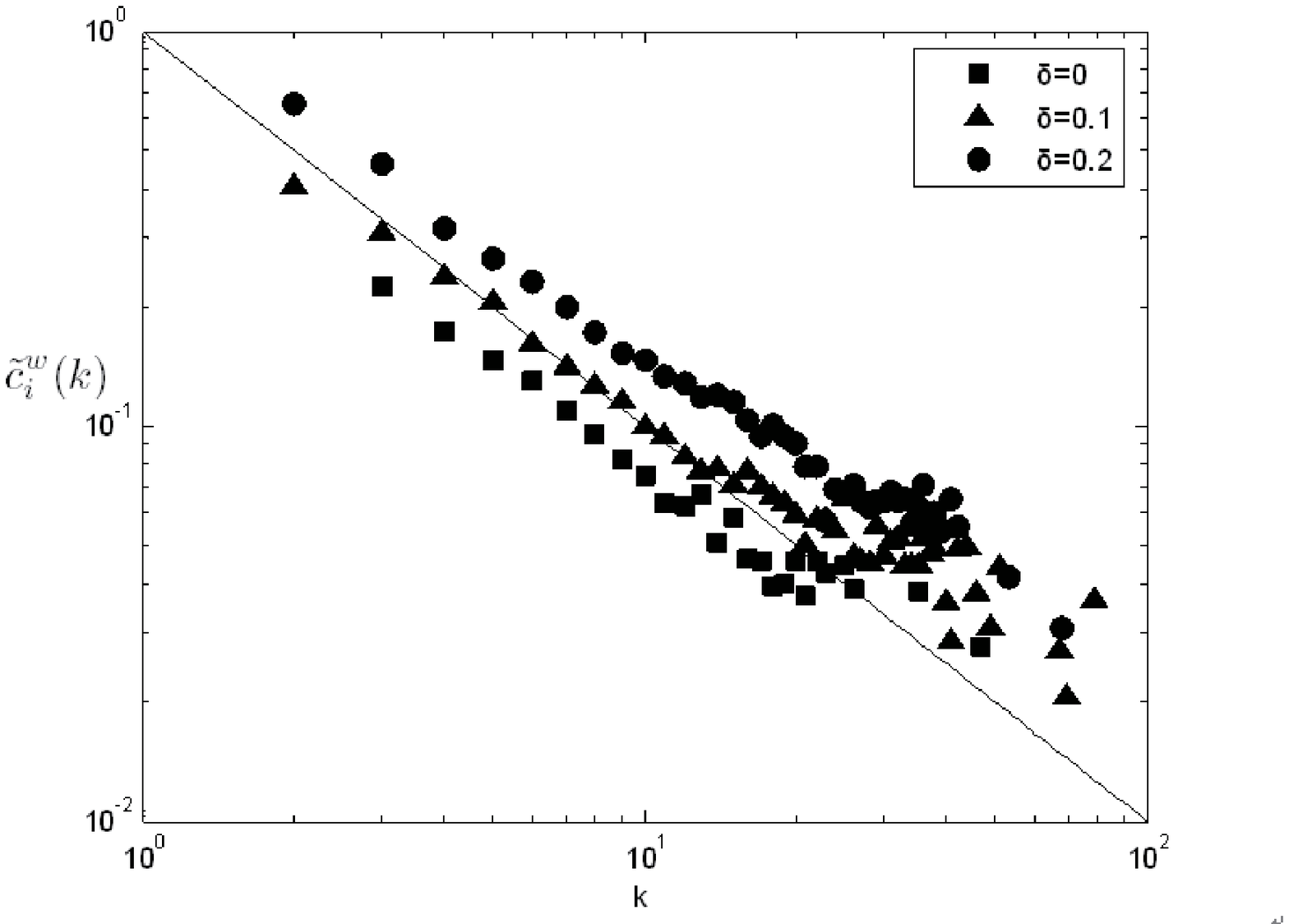}
\caption{ The weighted clustering coefficient $\tilde c_i^w(k)$ as a
function of the node degree $k$ for different $\delta$ in a log-log
plot with $m=3, p=0.1$ and $q=0.01$. Each data point is obtained by
averaging over 20 independent realizations with network size
$N=10^4$ fixed. The straight line of slope $-1$ is evident with
different data in plot.} \label{Fig:C3}
\end{figure}

{\em Case 2.} $-2A+1>0$. Then
$\frac{N^{-2A+1}-1}{-2A+1}\approx\frac{((qn_0+1-q)t)^{-2A+1}}{-2A+1}$
and $\sum_{j\in \Gamma_i}s_j(t)\approx
U\frac{((qn_0+1-q)t)^{-2A+1}}{2A-1}s_{i}(t)$. 
Hence 
\begin{equation} \label{eq:D15}
\begin{split}
\Delta_2&=\int^{t_{N}}_{t_{i}}m(m-1)(1+p)^2\frac{U(qn_0+1-q)^{-2A+1}}{2A-1}\frac{s_i^{2}t^{-2A+1}}{(\sum{N_{i})^{2}}\langle
s\rangle^2}\partial t\\
&=E_2\int^{t_N}_{t_i}\frac{(m(1+p))^2t^{-1}}{t_i^{2A}}\partial t\\
&=\frac{E_2(m(1+p))^2}{t_{i}^{2A}}(\ln t_N-\ln t_i),\\
\end{split}
\end{equation}
where $E_2=\frac{m(m-1)(1+p)^2U}{\langle
s\rangle^{2}(qn_{0}+1-q)^{2}}\frac{(qn_0+1-q)^{-2A+1}}{2A-1}$.
Ignoring the constant, the weighted clustering coefficient of $v_i$
with large strength $s_i(t_N)$ can be given by
\begin{equation} \label{eq:D16}
\begin{split}
&\tilde{c}_i^w=\frac{\langle w_{ijh}\rangle}{s_i(t_N)(k_i(t_N)-1)}(\Delta_1+\Delta_2)\\
&\approx \frac{\langle
w_{ijh}\rangle}{s_i(t_N)(k_i(t_N)-1)}(\frac{2mp}{A\langle
s\rangle(qn_{0}+1-q)}s_i(t_{N})+E_2s_i^2(t_n)\frac{\ln t_N}{t_N^{2A}}\\
&\approx \frac{2mp\langle w_{ijh}\rangle}{C\langle
s\rangle(qn_0+1-q)}\frac{1}{s_i(t_N)}+\frac{AE_2\langle w_{ijh}\rangle
}{C}\frac{\ln t_N}{t_N^{2A}}.
\end{split}
\end{equation}
In our model, the
parameter $p$ 
effects the system by allowing the formation of triads, the value
$\delta$ also affects $\tilde{c}_i^w$ without changing its power-law
slope. From Fig. \ref{Fig:C3}, we can find the weighted clustering
coefficient of each node can be adjusted continuously and growing
monotonically with an increasing $\delta$. In the expression of
$\tilde{c}_i^w$ of {\em Cases 1} and {\em  2}, the first term can be
viewed as the triad formation induced clustering, and it shows the
$s^{-1}$ behavior that has been observed in several real-life
systems \cite{Guimera1, Guimera2}.

{\em Case 3.} $-2A+1=0$.
 Then $q=0$ and $\delta=0$ in weighted tunable cluster local-world
network model, this case is in
accordance with  
the classic model produced by $P.Holme$ \cite{P.Holme}.
 Therefore, the weighted clustering coefficient for a node $v_i$ with
large degree $k_i(t_N)$ is
\begin{equation} \label{eq:D17}
\begin{split}
\tilde{c}_i^w=\frac{e_i}{k_i(k_i-1)/2}
\approx \frac{4p}{1+p}\frac{1}{k_i} +(1+p)^2\frac{m-1}{4(1+p)}\frac{(\ln N)^2}{N}.\\
\end{split}
\end{equation}
In (\ref{eq:D17}), the first term can be ascribed to the triad
formation which induced by clustering and displayed $k^{-1}$
behavior. 
We set $p$ any value in interval $(0,1)$, then $\tilde{c}_i^w$ can
be adjusted continuously and it will grow monotonically increasing
with $p$.   In this case, the model may be regarded as a growing
network with uniform attachment, which is similar to the triad
formation model proposed by Szab\'{o} et al.~\cite{szabo2003}.
Therefore Eq.~(\ref{eq:D17}) is similar to the expression of the
clustering coefficient in \cite{szabo2003}.

\subsection{Comparison with SCN Networks}

In order to verify the effectiveness of our in weighted tunable
cluster local-world network model, some real data of typical social
 networks are applied to our model in this section. In scientific co-authorship network
{\em (SCN)} of {\em MathSciNet}, nodes are defined as scientists and
two scientists (nodes) are connected if they have coauthored at
least one paper. Furthermore, we define the link's weight between
each pair of authors in an article with  $n$ authors by $1/(N-1)$.
Specially, our model can be applied to boolean networks if
 $\delta=0$.
We initialize our model with $c_0=3, n_0=3, p=0.05, q=0.1, m=4$,
when $T\geq 29981$, a boolean network with the same number nodes as
$SCN$ in 2004 \cite{angel} is obtained. As listed in Table 1, with
the same number of nodes $N$ and very similar number of edges $E$,
our model produces the same degree scaling exponent$\gamma$ with
boolean $SCN$ networks and produces very similar average
clustering coefficient $c$ and local worlds number $L$ and modularity $Q$. \\

\begin{tabular}{cc|c|c|c|c|c} 
\multicolumn{7}{c} {\bfseries Table 1. Topological properties of our
model and boolean SCN(2004)}
\\ \hline &\itshape N
&\itshape E &\itshape$\gamma$ &\itshape$c$ &\itshape
L &\itshape  Q
\\ \hline SCN(2004) & 30561 & 125959 &
3& 0.63 & 1069 & 0.668\\ \hline Model & 30561 & 115387 & 3& 0.6 & 1172 & 0.682\\
\hline Relative\,\,error & 0 & 0.08 & 0 & 0.04 & 0.09 & 0.02\\
\hline\label{tab1}
\end{tabular}

\begin{figure}[h!]
\centering
\includegraphics[angle=0, width=0.32\textwidth]{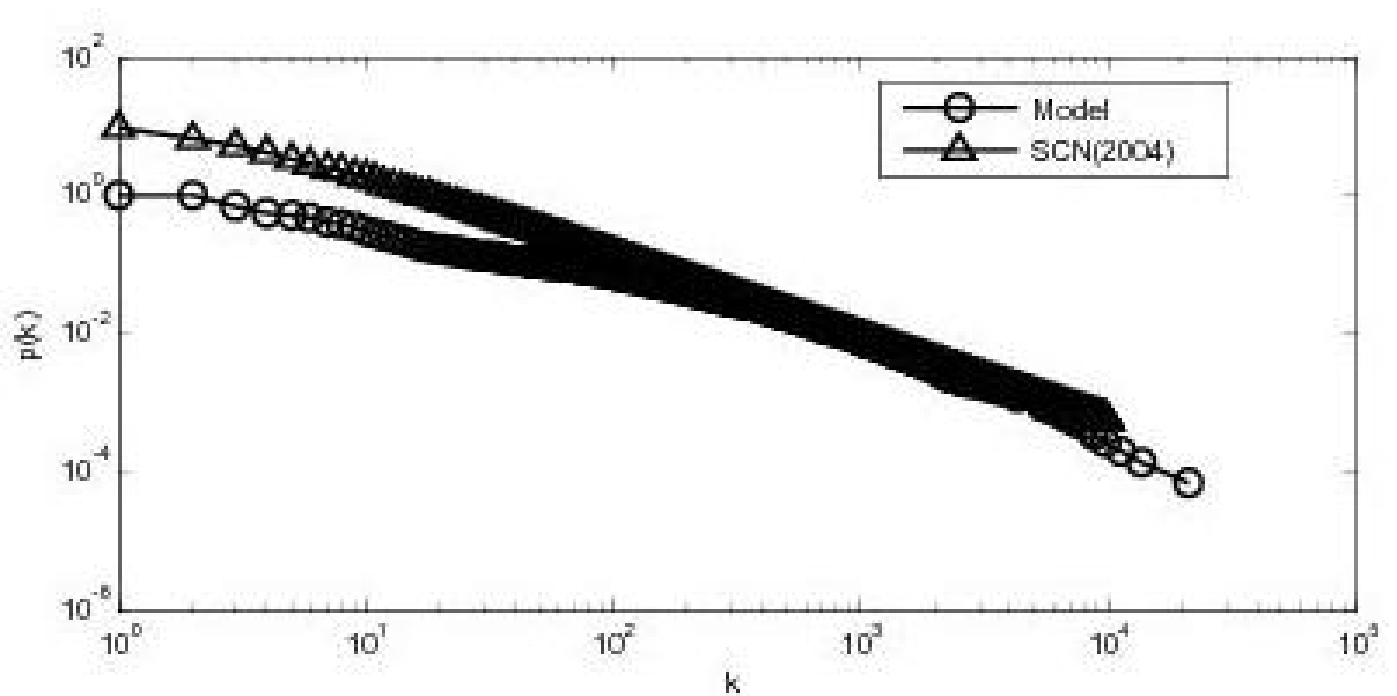}
\includegraphics[angle=0, width=0.32\textwidth]{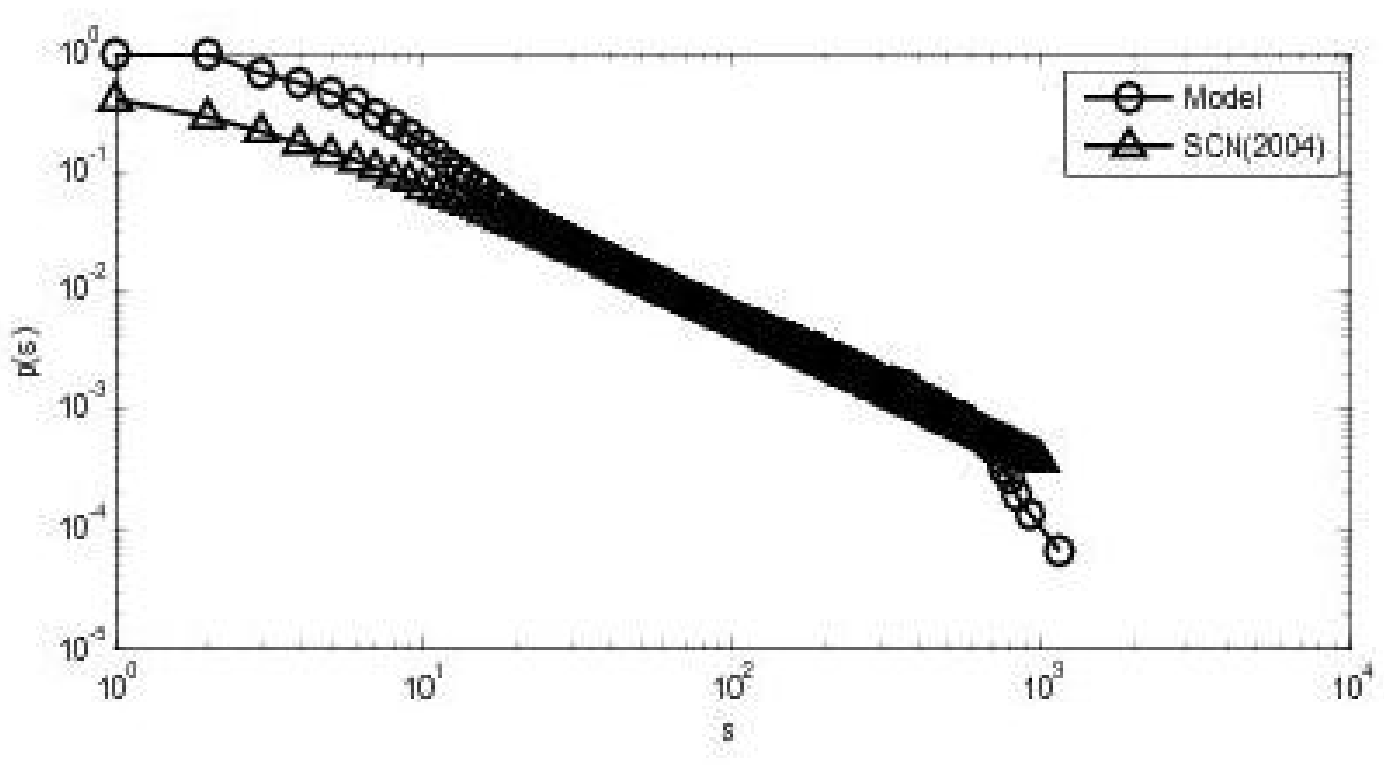}
\includegraphics[angle=0, width=0.32\textwidth]{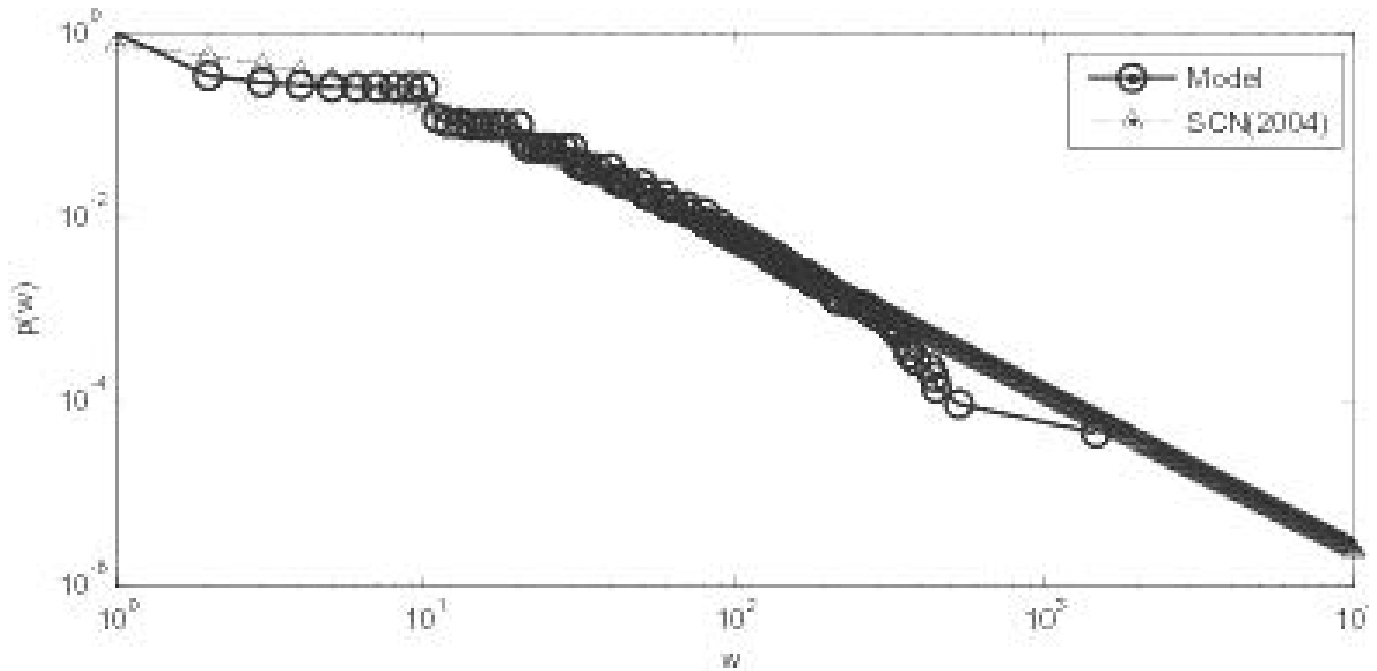}
\caption{Comparisons of our model and $SCN(2004)$ by cumulative distributions, (a) degree distribution $P(k)$,
(b) strength distribution $P(s)$, (c) weight distribution $P(w)$.} \label{Fig:10123}
\end{figure}

By cumulative distributions, the following figures shown that our
model owns very close degree distribution, strength distribution and
weight distribution compared to $SCN(2004)$ in Fig.\ref{Fig:10123},
the curves in $SCN(2004)$ are almost completely overlapped. In this
way, the network generated by our model matches to $SCN 2004$ well.

On  the other hand, our model is also compared with weighted
$SCN(2008)$. We initialized our model with $c_0=3,n_0=3,
p=004,q=0.09, m=3$, when $T\geq 9244$, a weighted network with the
same number of nodes as $SCN(2008)$ is obtained. Similarly, our
model is succeeded in capturing the structure properties of strength
scaling exponent $\gamma$, average clustering coefficient $c$ and
local world number $L$ and modularity $Q$ with the same number of
nodes $N$ and
very similar number of edges $E$, the results are shown in Table 2.\\

\begin{tabular}{lc|c|c|c|c|c}
\multicolumn{7}{c} {\bfseries Table 2. Topological properties of our
model and weighted SCN(2008)} \\ \hline &\itshape N
&\itshape E &\itshape$\gamma$ &\itshape$c$
&\itshape$n$ &\itshape$Q$\\ \hline SCN(2008) & 10136 & 31174 &
3.1& 0.65 & 254 & 0.679\\ \hline Model & 10136 & 30269 & 3.1& 0.64 & 231 & 0.682\\
\hline Relative\,\,error & 0 & 0.03 & 0 & 0.015 & 0.09 & 0.004\\
\hline\label{tab2}
\end{tabular}

\begin{figure}[h!]
\centering
\includegraphics[angle=0, width=0.3\textwidth]{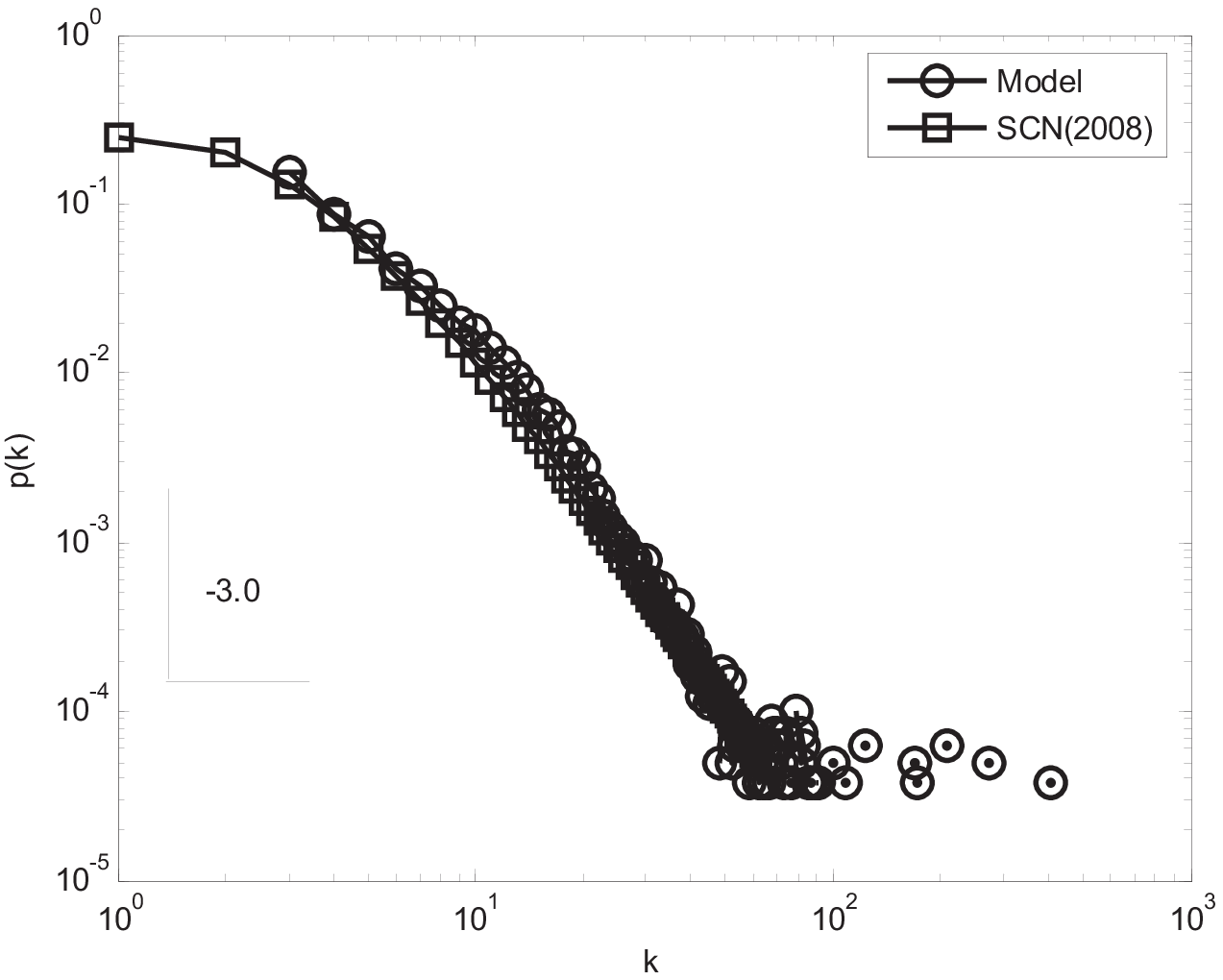}
\includegraphics[angle=0, width=0.3\textwidth]{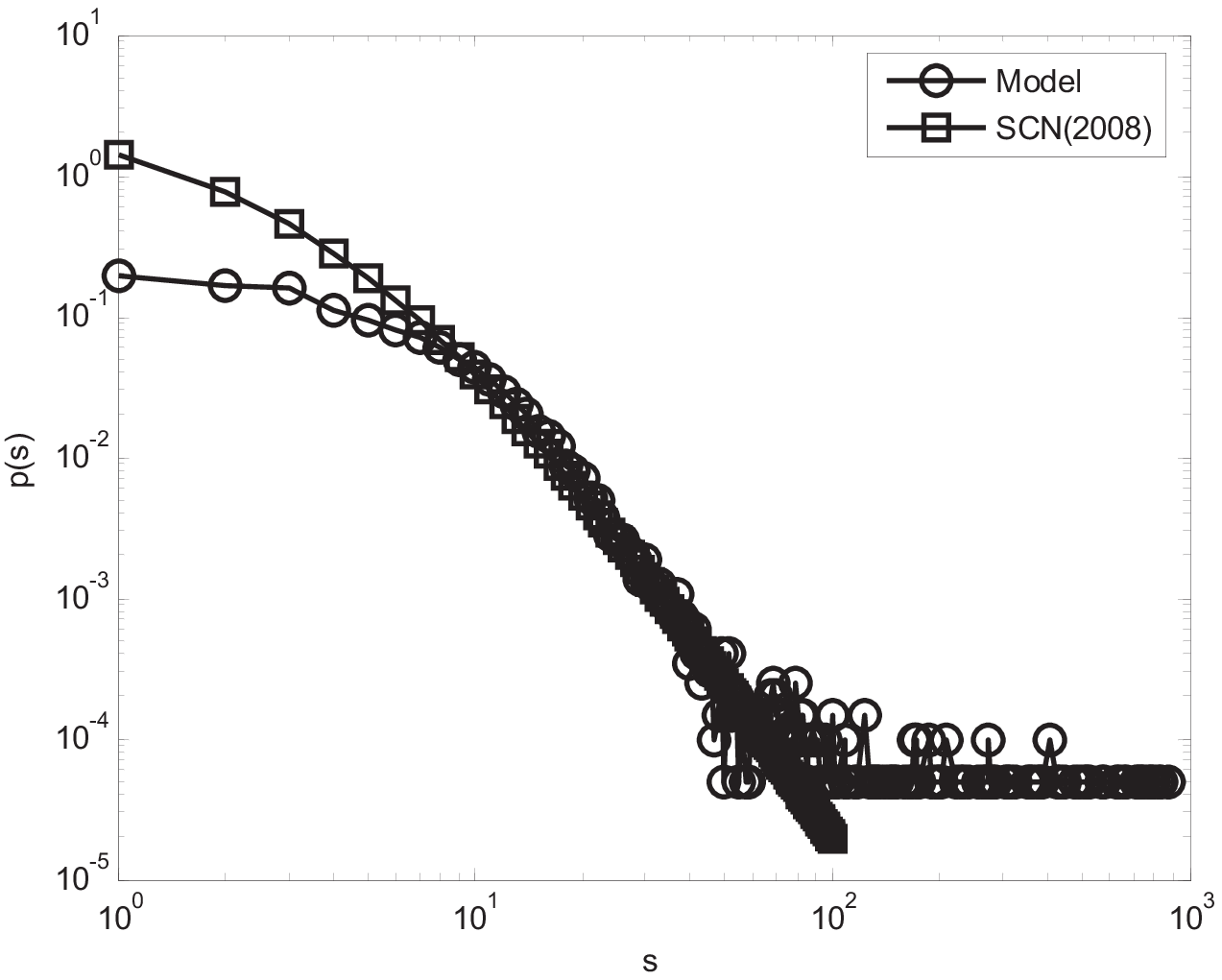}
\includegraphics[angle=0, width=0.3\textwidth]{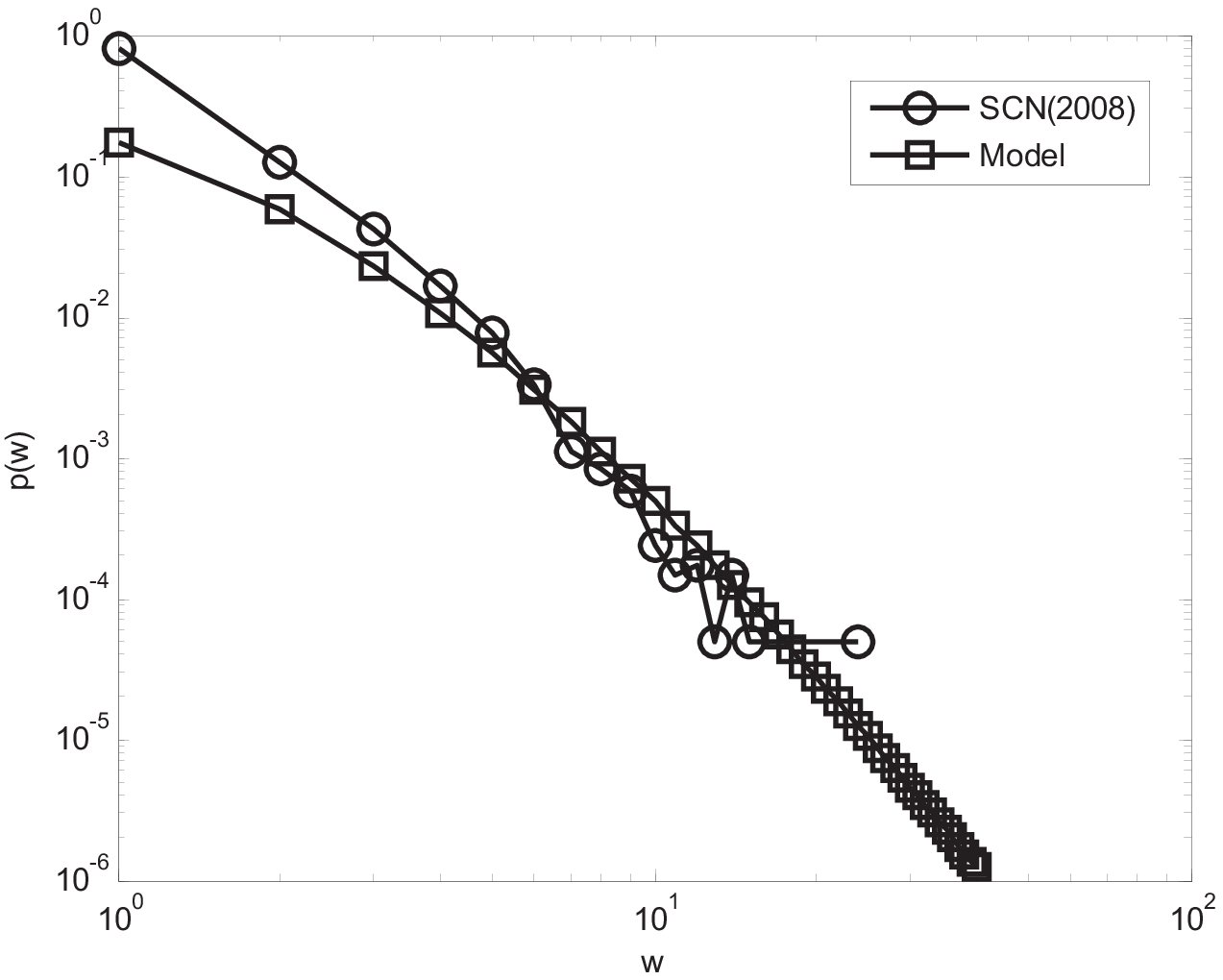}
\caption{Comparisons of our model and $SCN(2008)$,
(a) degree distribution $P(k)$, (b) strength distribution $P(s)$,
 (c) weight distribution $P(w)$.} \label{Fig:10456}
\end{figure}
Similarly, our model fits well the scale-invariant
 properties of degree distribution, strength distribution and weight distribution of the $SCN(2008)$, as shown in  Fig.\ref{Fig:10456}£®

\section{The correlation of vertices in model}

\begin{figure}[h]
\centering
\includegraphics[angle=0, width=0.4\textwidth]{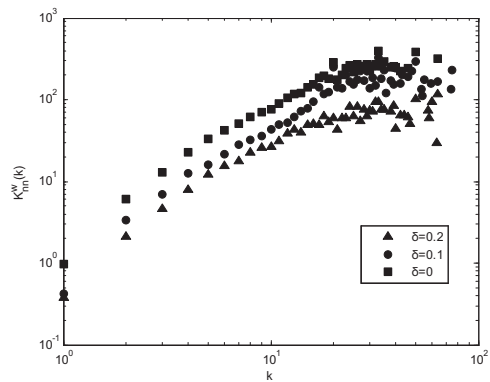}
\caption{The weighted degree-dependent average nearest-neighbor with
degree $k$, $k_{nn}^w(k)$, is a function of degree $k$ for different
$\delta$ in a log-log plot with $m=3, p=0.1$ and $q=0.01$. Each data
point is obtained over 20 independent average realizations with
network size $N=10^4$.} \label{Fig:C4}
\end{figure}

In the following section, we analyze the effect of degree
correlation between two nodes.  In real-world weighted networks, a
high degree vertex connects to low degree vertices with small
weight, while connects to high degree vertices with large weight.
For instance, in $WAN$, a busy airport $v_i$ has a lot of direct
flights to another dense airport $v_j$, while has a few of flight to
a spare airport.  In order to describe those phenomena,  Barrat et
al.\cite{Barrat} proposed the weighted degree-dependent average
nearest-neighbor with degree $k$, that is,

\begin{equation} \label{eq:3}
\begin{split}
 &k_{nn}^w(k)=\frac{1}{|\sum{v_i}|}\sum_{d_{v_i}=k}
 \frac{1}{s_i}\sum_{j\in \Gamma_i }w_{ij}k_j.\\
\end{split}
\end{equation}
In Fig. \ref{Fig:C4}, $k_{nn}^w(k)$ exhibits increasing power-law
behavior for $\delta\geq 0$ which indicated that the networks are
assortative. It is worth noting that a link introduced at an early
time, which connects two old vertices having similar degrees
together, tends to be high-weighted as $\delta\geq 0$. This leads to
the observed weighted assortative behavior.

Although we can find a network is assortative or disassortative by
$k_{nn}^w(k)$ increasing or decreasing with $k$, a quantity directly
describing the weighted assortatively is needed, it is introduced by
Barrat et al.\cite{Barrat}.

\begin{equation} \label{eq:D18}
\begin{split}
 &r^w=\frac{H^{-1}\sum_{\phi}(\varpi_{\phi}
 \prod_{i\in F(\phi)}k_i)-[\frac{H^{-1}}{2}
 \sum_{\phi}(\varpi_{\phi}\prod_{i\in F(\phi)}k_i)]^2}{\frac{H^{-1}}{2}
 \sum_{\phi}(\varpi_{\phi}\prod_{i\in F(\phi)}k_i^2)-
 [\frac{H^{-1}}{2}\sum_{\phi}(\varpi_{\phi}\prod_{i\in F(\phi)}k_i)]^2}.\\
\end{split}
\end{equation}
where $\varpi_{i}$ is the weight of the $\phi th$ link, $F(\phi)$ is
the set of the two vertices connected by the $\phi-th$ link and $H$
is the total weight of all links in the network. $r^w$ lies between
-1 and 1. Moreover, $r^w$ is positive for weighted assortative
networks,  is negative for otherwise.

\begin{figure}[h]
\centering
\includegraphics[angle=0, width=0.4\textwidth]{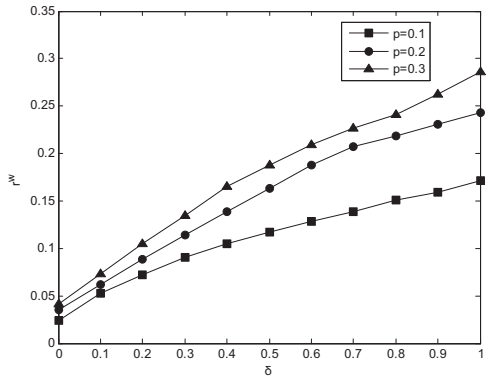}
\caption{The weighted assortativity coefficient $r^w$ as a function
of the node degree $\delta$ for different $p$ with $m=3, q=0.01$.
Each data point is obtained by averaging over 20 independent
realizations with network size $N=10^4$ fixed.} \label{Fig:C5}
\end{figure}

In Fig.~\ref{Fig:C5}, it is shown that $r^w$ is positive and
increases with the growth of $\delta$ and for all $q$. Thus, the
networks are assortativity for all $\delta$, which is in accord with
the definition of $k_{nn}^w(k)$. Therefore, the probability $p$ of
adding new $TF$ links and the value of $\delta$ significantly
determine the degree-degree mixing patterns of evolving weighted
networks in our model. This phenomenon can be easily understood
since the local-world inherits the property of the hierarchy
structure in networks.

\section{Epidemic spreading in our model with weighted transmission rate}

Epidemic spreading theory on complex networks  has a board practical
background. To study the dynamics of infectious diseases spreading
on the weighted tunable cluster local-world with increment behavior
network model  making the transmission rate accord with the
realistic cases much more, we take into account the effects of the
weights of edges and the strengths of nodes which are of great
importance measures in the weighted networks \cite{M.E.J.
Newman}-\cite{Guimera1}. For epidemic spreading, the weight can
indicate the extent of frequency of the contacting of two nodes in
scale-free networks, the larger the weight is, the more intensively
the two nodes communicate, the more possible a susceptible
individual will be infected through the edge, where the transmission
rate is larger.

 We make use of the spreading related weight $w_{kk'}$ between two nodes with
 degree $k$ and $k'$ represented as a function of their degrees
 \cite{M.E.J. Newman}-\cite{serrano},
$w_{kk'} = w_0(kk')^\beta$, where the basic parameter $w_0$ and the
exponent $\beta$ depend on the particular complex networks (e.g., in
the {\em E.coli} metabolic network $\beta = 0.5$; in the US airport
network (USAN) $\beta = 0.9$ \cite{Newman M E J}; in the scientist
collaboration networks (SCN) $\beta = 0$ \cite{M.E.J. Newman}).
Noteworthily, the spreading related weight $w_{kk'}$ belongs to an
edge which connect a  node of degree $k$  and a node of degree
$k^{\prime}$, the spreading related weights of links connected to
all nodes of  degree $k$  is $s_k = k\sum_{k'}P(k'/k)w_{kk'} $, and
$s_k$ is also the strength of a node of degree $k$. In the
following, we focus on uncorrelated (also called non-assortative
mixing) networks where the conditional probability satisfies
$P(k'/k) = k'P(k')/\langle k_i\rangle$ \cite{J.D. Murray}. Thus, one
can obtain $s_k = w_0 \langle k^{1+\beta}\rangle k^{1+\beta}/\langle
k\rangle$.

Here, the total transmission rate of each $k$-degree node is
$\lambda k$, since the transmission is $\lambda$ for each link
adjacent to this node. A transmission rate on the edge from the
$k$-degree node to $k'$-degree node, say $\lambda_{kk'}$, is defined
as follows \cite{Barrat},

\begin{equation} \label{eq:E1}
\begin{split}
&\lambda_{kk'} = \lambda k\frac{w_{kk'}}{s_k}.
\end{split}
\end{equation}
In (\ref{eq:E1}), the more proportion of $s_k$ that $w_{kk'}$ of an
edge holds, the more possible the disease will transmit through this
edge.   $\lambda_{kk'}=\lambda k^\beta \langle k\rangle \langle
k^{1+\beta}\rangle$ in uncorrelated networks.

In the following, we investigate the modified $SI$ model \cite{J.D.
Murray} in which the weighted transmission rate and nonlinear
infectivity are introduced. The results we obtain might deliver some
useful information for the epidemic outbreak. And for a better
analysis, we firstly describe the general differential equations for
$SI$ model based on the mean field theory, as follows:

\begin{equation} \label{eq:E2}
\begin{split}
&\partial I_{k}(t)/\partial t=
k(1-I_{k}(t))\sum_{k'}P(k'/k)I_{k'}(t)\lambda_{kk'},
\end{split}
\end{equation}
where $I_{k}$ and $\lambda_{kk'}$ denote the $k$-degree nodes'
infectivity and the transmission rate from $k$-degree nodes to
$k'$-degree nodes respectively. Neglecting the terms of $O(k^2)$ in
the expansion of (\ref{eq:E2}), the simplified result is

\begin{equation} \label{eq:E3}
\begin{split}
\partial I_{k}(t)/\partial t&=\frac{\lambda k^{1+\beta}\langle
k\rangle}{\langle k^{1+\beta}\rangle}\sum_{k'}\frac{k'P(k')}{\langle
k\rangle}I_{k'}(t) =\frac{\lambda k^{1+\beta}\langle
k\rangle}{\langle k^{1+\beta}\rangle}\theta_k(t).
\end{split}
\end{equation}
In uncorrelated heterogeneous networks, $\theta_k(t)$ is independent
of the degree of vertex, then
$\theta_k(t)=\theta(t)=\sum_k\frac{kP(k)}{\langle k\rangle}I_k(t)$
and every infected neighbor may be the initial seeds(infected at
$t=0$) or be infected at $t>0$.  Therefore,

\begin{equation} \label{eq:E6}
\begin{split}
I_{k}(t)=I_{0}+I_{0}\frac{\lambda k^{1+\beta}\langle
k\rangle}{\langle k^{1+\beta}\rangle}(e^{t/\tau}-1),
\end{split}
\end{equation}
where $\tau=\frac{\langle k^{1+\beta}\rangle}{\lambda \langle
k^{2+\beta}\rangle}$, and the total infection density is

\begin{equation} \label{eq:E7}
\begin{split}
I(t)=\sum_{k}P(k)I_{k}(t)=I_{0}+I_{0}\frac{\lambda \langle
k^{1+\beta}\rangle\langle k\rangle}{\langle
k^{1+\beta}\rangle}(e^{t/\tau}-1),
\end{split}
\end{equation}
since $\partial \theta(t)/\partial t=\sum_{k}\frac{kP(k)}{\langle
k\rangle}\frac{\partial I_{k}(t)}{\partial t}=\frac{\lambda \langle
k^{2+\beta}\rangle}{\langle k^{1+\beta}\rangle}\theta(t)$ and the
uniform initial condition $I_k(t=0)=I_{0}$.

In the weighted tunable cluster local-world with increment behavior
network model, the probability density of $k$-degree node is
$p(k)=ak^{-1-1/A}$ and this model is also an uncorrelated networks,
$\tau$ is obtained in the following
\begin{equation} \label{eq:E8}
\begin{split}
\tau=\frac{\langle k^{1+\beta}\rangle}{\lambda \langle
k^{2+\beta}\rangle}=\frac{\int^{k_{max}}_1k^{\beta+1}p(k)\partial
k}{\int^{k_{max}}_1k^{\beta+2}p(k)\partial k}
=\frac{2+\beta-\frac{1}{A}}{\lambda(1+\beta-\frac{1}{A})}\cdot
\frac{k_{max}^{\beta+1-\frac{1}{A}}-1}{k_{max}^{\beta+2-\frac{1}{A}}-1}.
\end{split}
\end{equation}
Generally, when $1+\beta-\frac{1}{A}>0$, namely
$2+\beta>\gamma=1+\frac{1}{A}$. If $k_{max}$ big enough, one can get

\begin{equation} \label{eq:E9}
\begin{split}
&\tau\approx\frac{2+\beta-\frac{1}{A}}{\lambda(1+\beta-\frac{1}{A})}\cdot\frac{1}{k_{max}}.
\end{split}
\end{equation}
\begin{figure}[h]
\centering
\includegraphics[angle=0, width=0.32\textwidth]{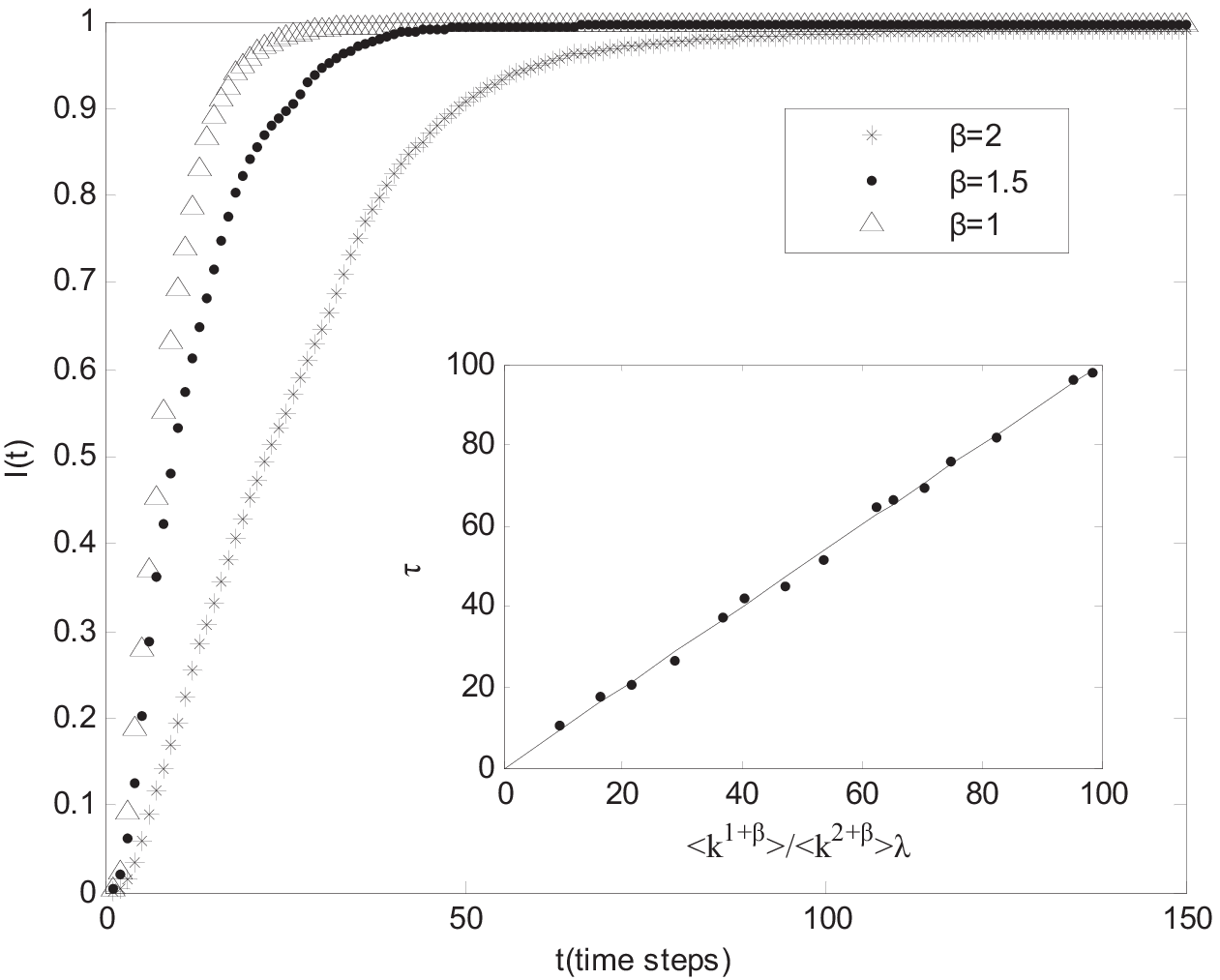}\,
\includegraphics[angle=0, width=0.32\textwidth]{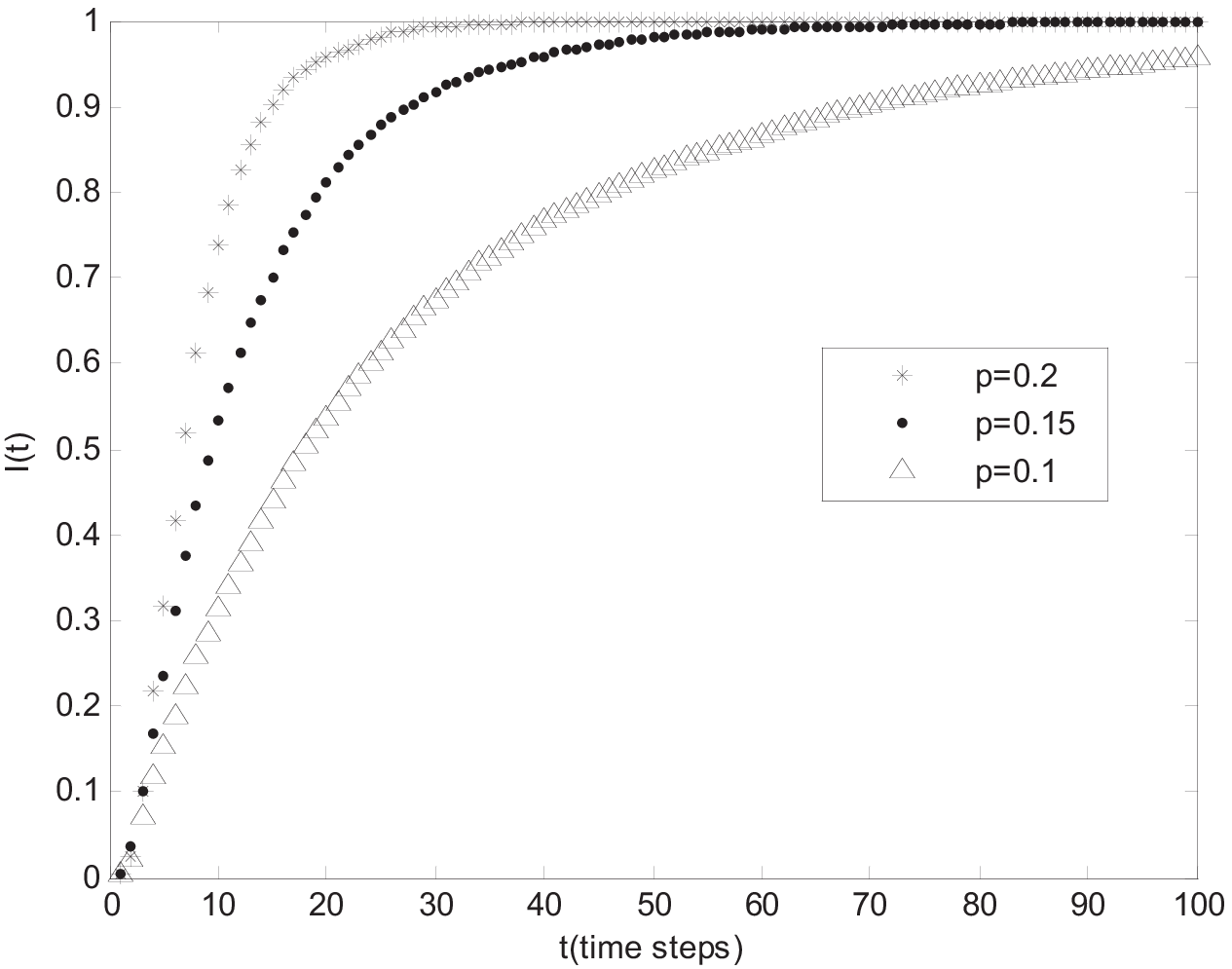}\,
\includegraphics[angle=0, width=0.32\textwidth]{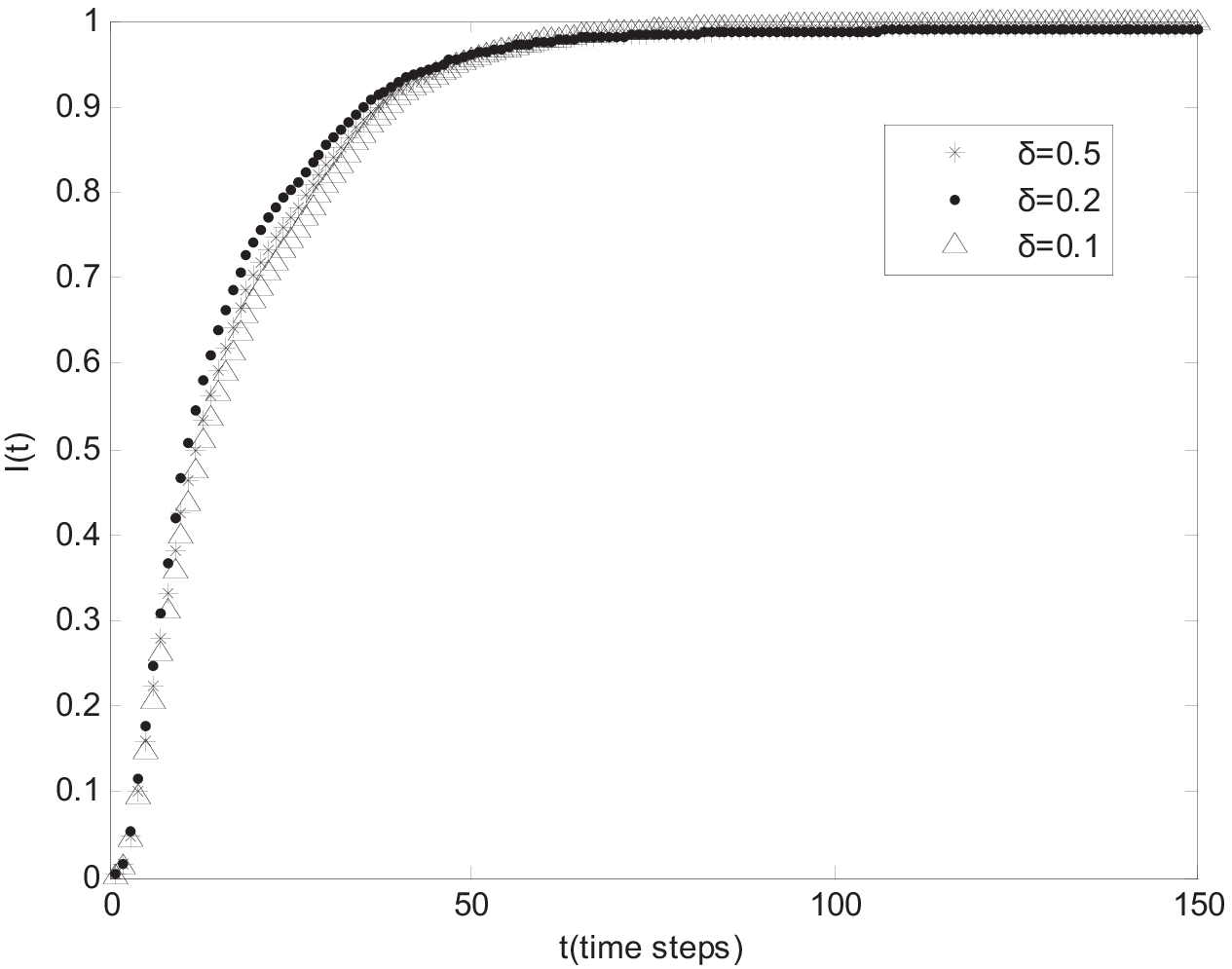}
\caption{The left figure shows the dynamical of the infection
density for different $\beta$ with $m=3, p=0.1,q=0.01$ and
$\delta$=0.1. The inner graph shows the relation of the time scale
$\tau$ with the theoretical result. The middle figure illustrates
the dynamical of the infection density for different $p$ with $m=3,
q=0.01, \beta=1$ and $\delta=0.1$. The right figure shows the
dynamical of the infection density for different $\delta$ with $m=3,
p=0.1, \beta=1$ and $q=0.01$. All the curves are obtained by
averaging over 50 independent realizations.} \label{Fig:C6}
\end{figure}
The parameter $A$ can be treated as a function of $p$, $q$ and
$\delta$, according to (\ref{eq:E7}), then $\tau$ and  $I(t)$
increase with the degree of tunable clustering mechanism and
inversely effect $\beta$. In order to illustrate this property, some
numerical simulations are shown in the left graph of
Fig.\ref{Fig:C6}, $\beta$ accelerates the dynamical evolution when
$\beta$ is less and the inner graph shown the time scale $\tau$
agrees well with the theoretical results.
 The dynamical evolution of infection density under
various value $p$ is explained by the middle picture of
Fig.\ref{Fig:C6}. The larger value of $p$, that is, the larger
probability of $TF$ links occurs, the fast the dynamical evolution
reaches steady state. In the right of Fig.\ref{Fig:C6} plots the
effect with different $\delta$ on the dynamic spreading behavior.
However, $\delta$ doesn't have obvious effect on the dynamic of
$I(t)$. A reasonable explanation is that the increment behaviors do
not change the value of $k$ nor $p(k)$, and therefor do not effect
the weighted transmission rate $\lambda_{kk'}$. The current analysis
and simulations provide an intuitive description of the spreading
phenomena in complex networks, and it helps us to further understand
some real-world propagation mechanisms and spreading behaviors.

\section{Conclusion}

Tunable clustering mechanism and the increment behavior are two
important dynamic mechanisms in real world networks. In order to
interpret
 the real network accurately, we  propose a weighted tunable cluster
 local-world evolving model with increment behavior.
 The growth dynamics include some new local-world and new nodes added by local
preferential attachment. A lot of interesting properties of the
generated networks display good right-skewed distribution
characters, which have been discovered in most realistic networks
such as the distributions of strength, degree and link weight
display of a power law. We also performed numerical simulations and
verified the experimental results  which are agreement with
theoretical analysis very well. The effects of tunable clustering
mechanism and increment behavior on correlations of vertices of
weighted networks are studied also. The accurate values of the
weighted clustering coefficient, weighted average nearest-neighbor
degree and weighted assortatively coefficient are obtained. All of
these results exhibit the assortative behaviors of our model and the
dynamic mechanisms properties. Finally, it is discovered that the
tunable clustering behavior has a great impact on the spreading
dynamic. However, due to the particularity of the epidemic spreading
model, the increment behavior doesn't have obvious effect on the
spreading dynamic in our model.

\begin{center}
{\bf Acknowledgements}
\end{center}

The authors would like to thank the  anonymous referees very much
for
 valuable suggestions, corrections and comments which results in  a great improvement of
the original manuscript. In particular, the
 referee provided  some suggestion on Fig. \ref{Fig:10123} and references
 \cite{onnela2005, holme2007, szabo2003}.


\end{document}